# Cluster Diffusing Shuffles

Kevin Su — May 2020

# Abstract

Unbiased shuffling algorithms, such as the Fisher-Yates shuffle, are often used for shuffle play in media players. These algorithms treat all items being shuffled equally regardless of how similar the items are to each other. While this may be desirable for many applications, this is problematic for shuffle play due to the **clustering illusion**, which is the tendency for humans to erroneously consider "streaks" or "clusters" that may arise from samplings of random distributions to be non-random.

This thesis attempts to address this issue with a family of biased shuffling algorithms called **cluster diffusing (CD) shuffles** which are based on disordered hyperuniform systems such as the distribution of cone cells in chicken eyes, the energy levels of heavy atomic nuclei, the eigenvalue distributions of various types of random matrices, and many others which appear in a variety of biological, chemical, physical, and mathematical settings. These systems suppress density fluctuations at large length scales without appearing ordered like lattices, making them ideal for shuffle play.

The CD shuffles range from a random matrix based shuffle which takes $O(n^3)$ time and $O(n^2)$ space to more efficient approximations which take $O(n)$ time and $O(n)$ space.

## Acknowledgements

Very special thanks to Dr. Lippold Haken, my thesis adviser, for his patience and willingness to support this thesis despite its very late start in the thesis registration timeline.

I would also like to thank Lukáš Poláček for kindly providing unreleased details about the Poláček Shuffle.

# Contents

# 1 Human Perception of Randomness

Unbiased shuffling algorithms, such as the Fisher-Yates shuffle, are often used for shuffle play in media players. These algorithms treat all items being shuffled equally regardless of how similar the items are to each other. While this may be desirable for many applications, this is problematic for shuffle play due to the **clustering illusion**, which is the tendency for humans to erroneously consider "streaks" or "clusters" that may arise from small samplings of random distributions to be non-random.[1]

For example, when asked to emulate a single, short series of fair coin tosses, people will tend to produce sequences with shorter streaks and higher alternation frequencies than expected.[2,3,4]

Another example is R.D. Clarke's analysis of the distribution of flying-bomb impacts across London during World War II; certain areas were hit more than others, leading to frequent assertions that the impact points tended to cluster (Figures 1, 2, 3).

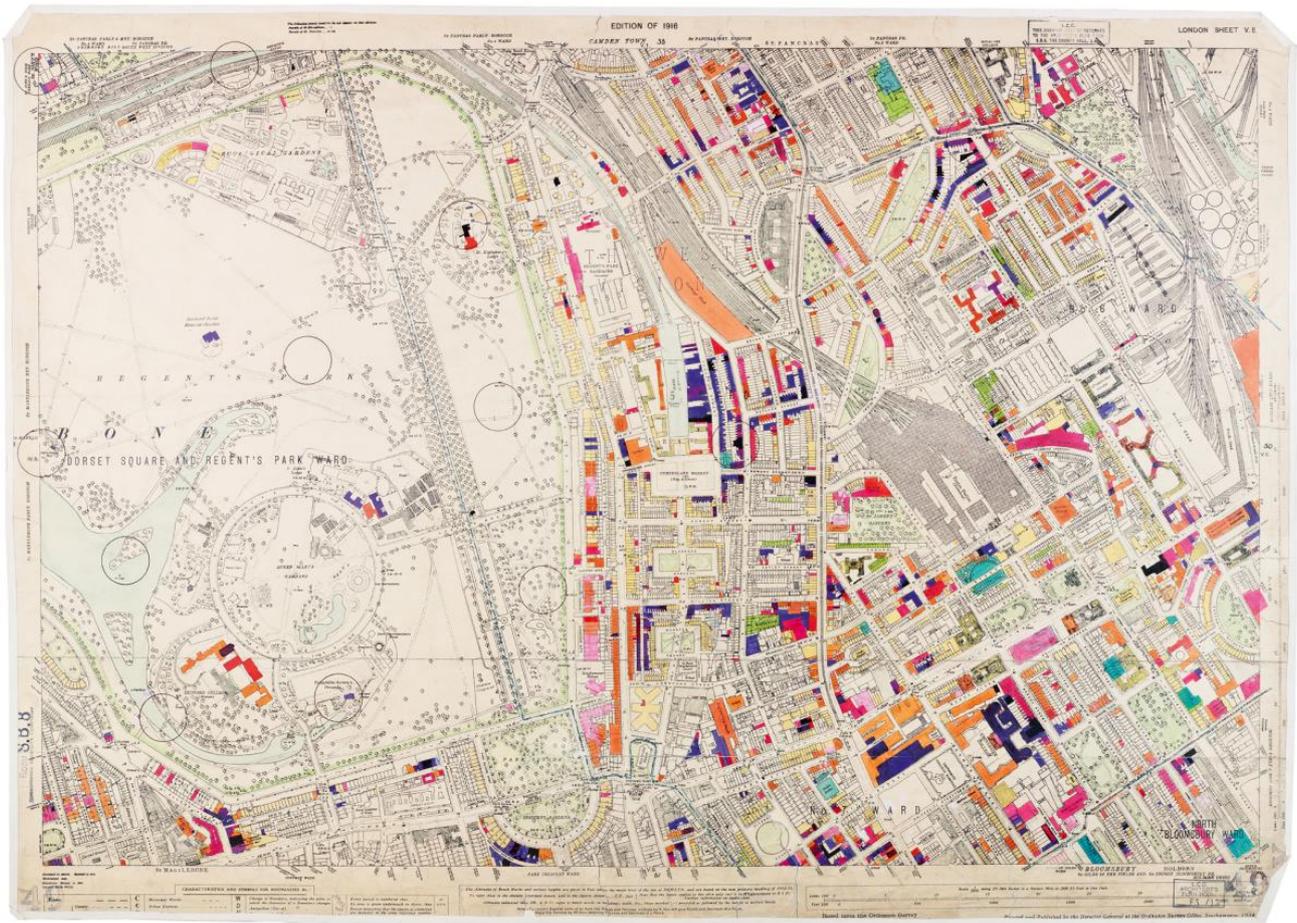

*Figure 1: London City Council Bomb Damage Map for Regent's Park and Somerstown.*[5]

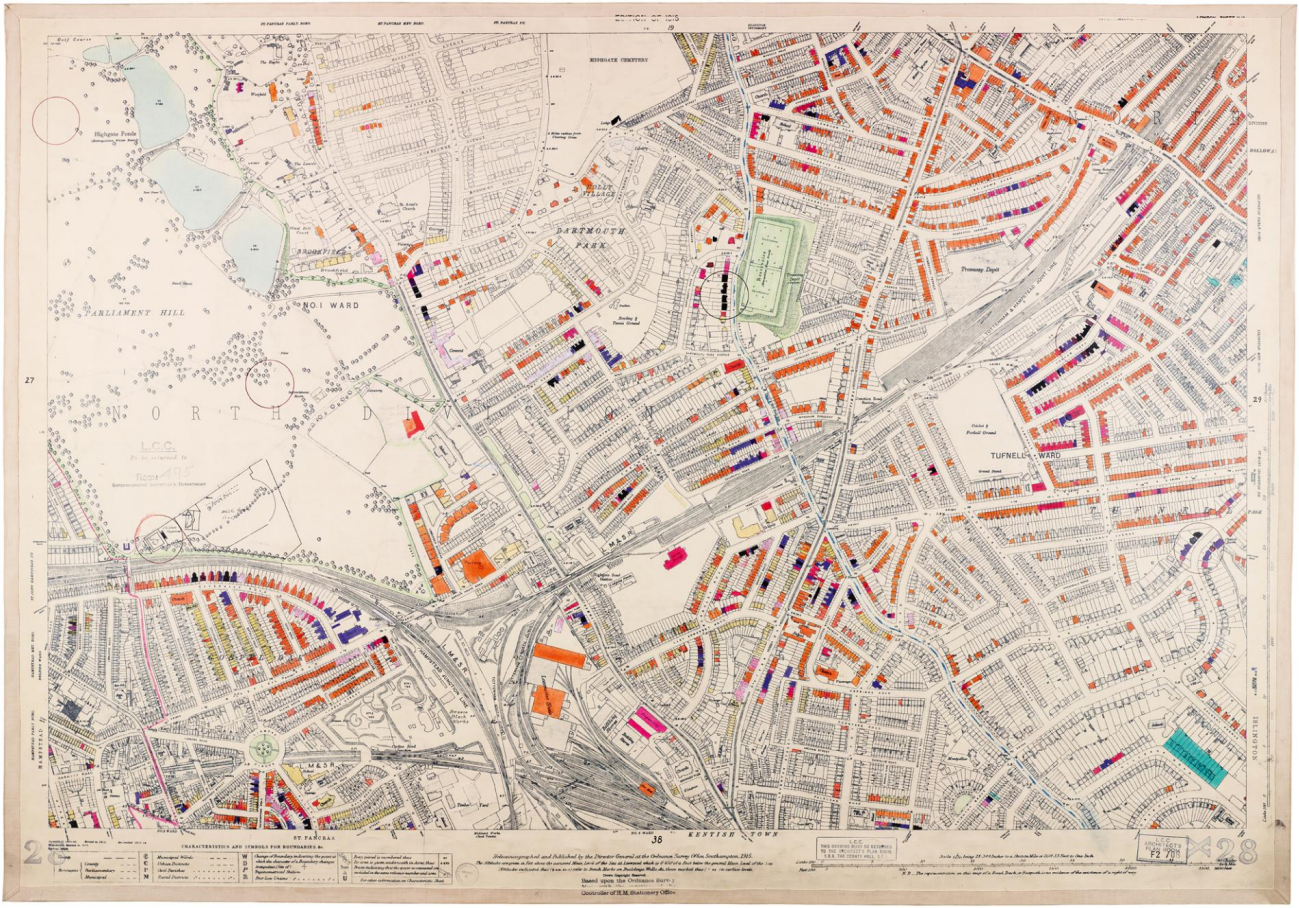

*Figure 2: London City Council Bomb Damage Map for Upper Holloway and Parliament Hill.[5]*

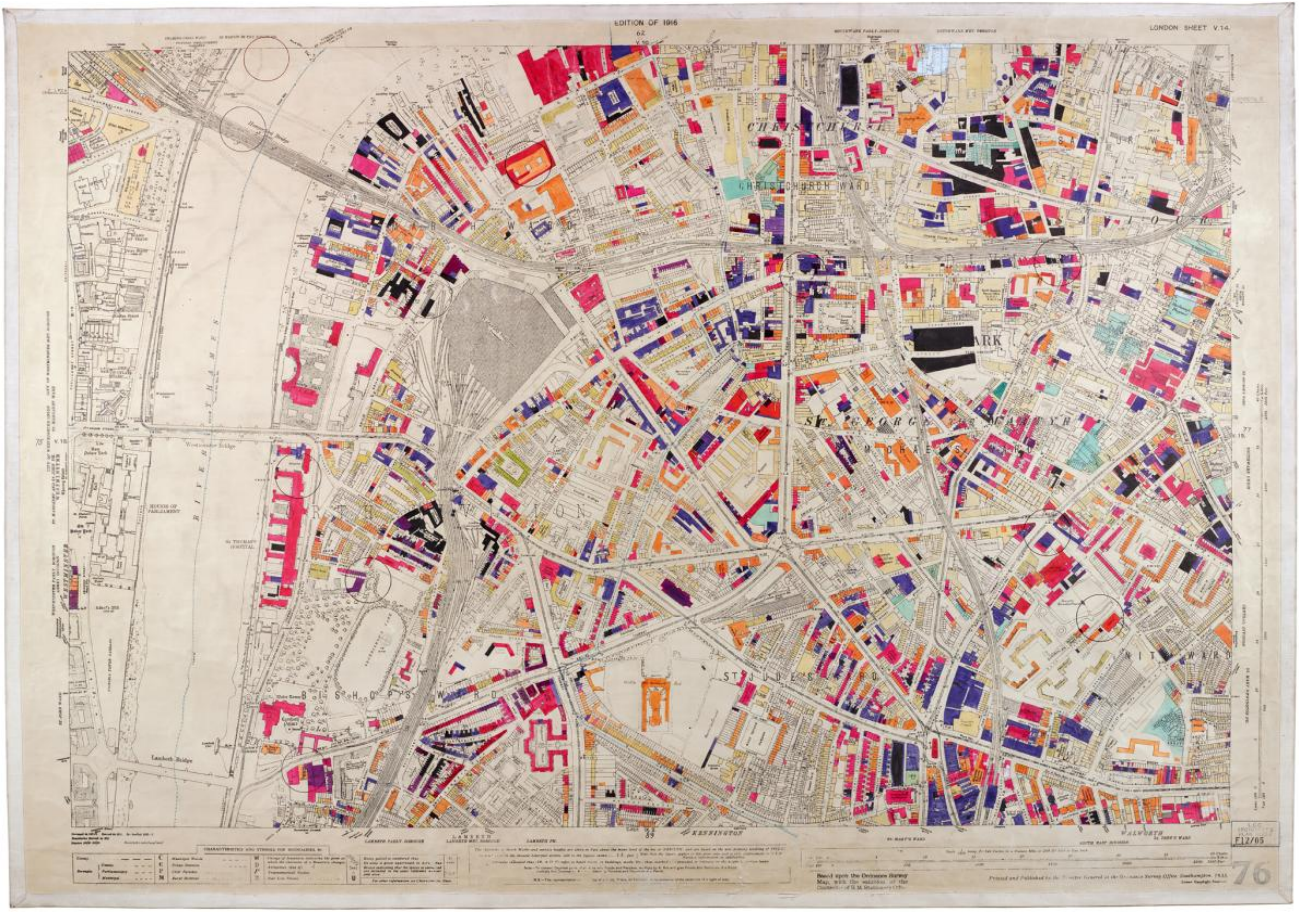

*Figure 3: London City Council Bomb Damage Map for Waterloo and Elephant & Castle.[5]*

Clarke's analysis, however, showed that the distribution of impact densities closely conformed to the Poisson distribution derived from a model that assumed completely random bombing (Table 1).[6]

| No. of flying bombs per square | Expected no. of squares (Poisson) | Actual no. of squares |
|---|---|---|
| 0 | 226·74 | 229 |
| 1 | 211·39 | 211 |
| 2 | 98·54 | 93 |
| 3 | 30·62 | 35 |
| 4 | 7·14 | 7 |
| 5 and over | 1·57 | 1 |
| | 576·00 | 576 |

Table 1: Comparison between expected and actual distributions of flying bomb impact densities for ¼ $km^2$ square plots.[6]

In the context of music, Spotify found that many of the complaints about their shuffle play algorithm, which was the Fisher-Yates shuffle at the time, being non-random were about songs from the same artist being tightly clustered within a small portion of the playlist instead of being evenly spread out (Figure 4).[7]

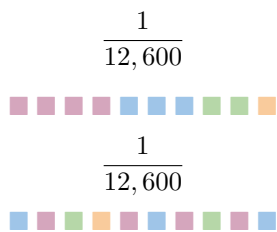

Figure 4: Two potential shuffles of songs (colored by artist) from an unbiased shuffling algorithm and their probabilities.

## 2   Assumptions

For the time and space complexity analyses, we will use the uniform cost model and assume that the machine is using a fixed, finite number of bits, allowing for operations such as radix sort to take $O(n)$ time and space.

In addition, sampling from any probability distribution is assumed to take $O(1)$ time and space per sample.

# 3 Reference Shuffles

## 3.1 Unbiased Shuffles

Consider an unbiased shuffle which shuffles items by assigning each item to a unique random number sampled from a uniform random variable (r.v.) on $[0, 1]$ and then sorting the items by their assigned number. The spacing distribution of any $n$ items is given by Equation 1, where $x$ is the distance between consecutive items (Figure 5).[8]

$$p_{uniform\ spacing}(x) = n(1-x)^{n-1} \tag{1}$$

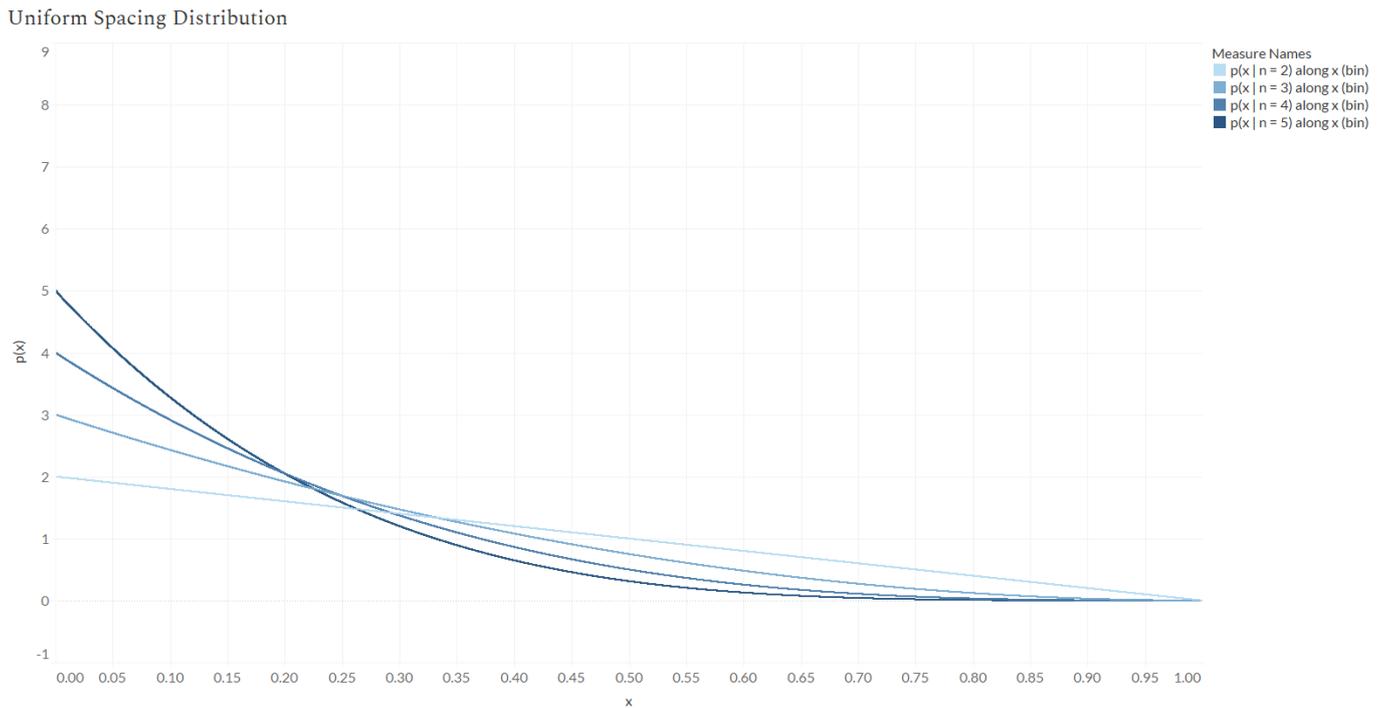

*Figure 5: Uniform spacing distributions for $n = [2, 3, 4, 5]$.*

Items are most likely to be assigned numbers that are close to each other than numbers that are far apart. Whether this is undesirable for items from different groups is undecided. If the items are from the same group, however, this is undesirable because they are more likely to cluster than not.

## 3.2 The Balanced Shuffle

Martin Fiedler attempted to address the clustering illusion problem with the Balanced Shuffle in 2007.[9] The Balanced Shuffle first shuffles each song group with an unbiased shuffle and then pads the groups with spacers to match the length of the longest group. Within each group, these spacers are inserted at positions such that each song resides in its own interval. These intervals partition the group and are of relatively or exactly even length depending on if the number of songs in the group evenly divides the size of the longest group (Figure 6).

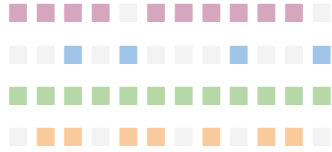

*Figure 6: An example output from the Balanced Shuffle's padding step.[9]*

Afterwards, each group is traversed simultaneously. During the traversal, the songs at the same index are assembled into a list. This resulting list is shuffled using an unbiased shuffle and then appended to the output playlist.

The padding step causes the Balanced Shuffle to have a potentially high time and space requirement of $O(n^2)$, such as on inputs where one group has many songs while the other groups only have a few songs each (Figure 7).

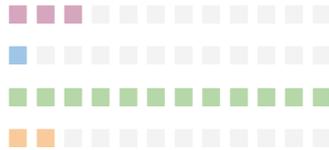

*Figure 7: An example of an inefficient input for the Balanced Shuffle.*

During the padding step, song groups whose number of songs evenly divides the size of the longest group will partition their padded selves into intervals of even length. Groups that don't, however, will partition themselves randomly into intervals of only approximately even length, making them difficult to analyze (Figures 8, 9).

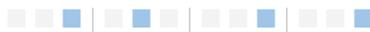

*Figure 8: A padded group from Figure 6 which is partitioned into **even** intervals.*

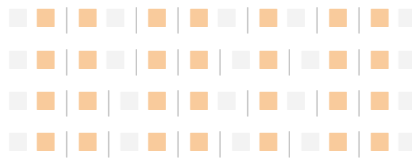

*Figure 9: A padded group from Figure 6 which is partitioned into **uneven** intervals.*

We can circumvent this by extending the Balanced Shuffle from the integers ($\mathbb{Z}$) to the real numbers ($\mathbb{R}$). When padding the song groups, we pretend that there is an extra group that has an infinite number of songs. This allows each song group to be partitioned into equi-width intervals and also lets us observe the song distribution behavior of the padding step without needing to account for the random, uneven partitioning side-effect that comes from the limited song and spacer placement granularity imposed by a longest group with finite length.

Since songs within a group can occupy any position within their assigned intervals, we can model song positions as a set of equi-width uniform r.v.s that partition $[0, 1]$. Thus, the spacing distribution of a song group is given by Equation 2, where $\Lambda(x)$ is the unit triangle function and $n$ is the number of songs in the group (Figure 10).

$$p_{Balanced\ spacing}(x) = n\Lambda(n(x - \frac{1}{n})) \tag{2}$$

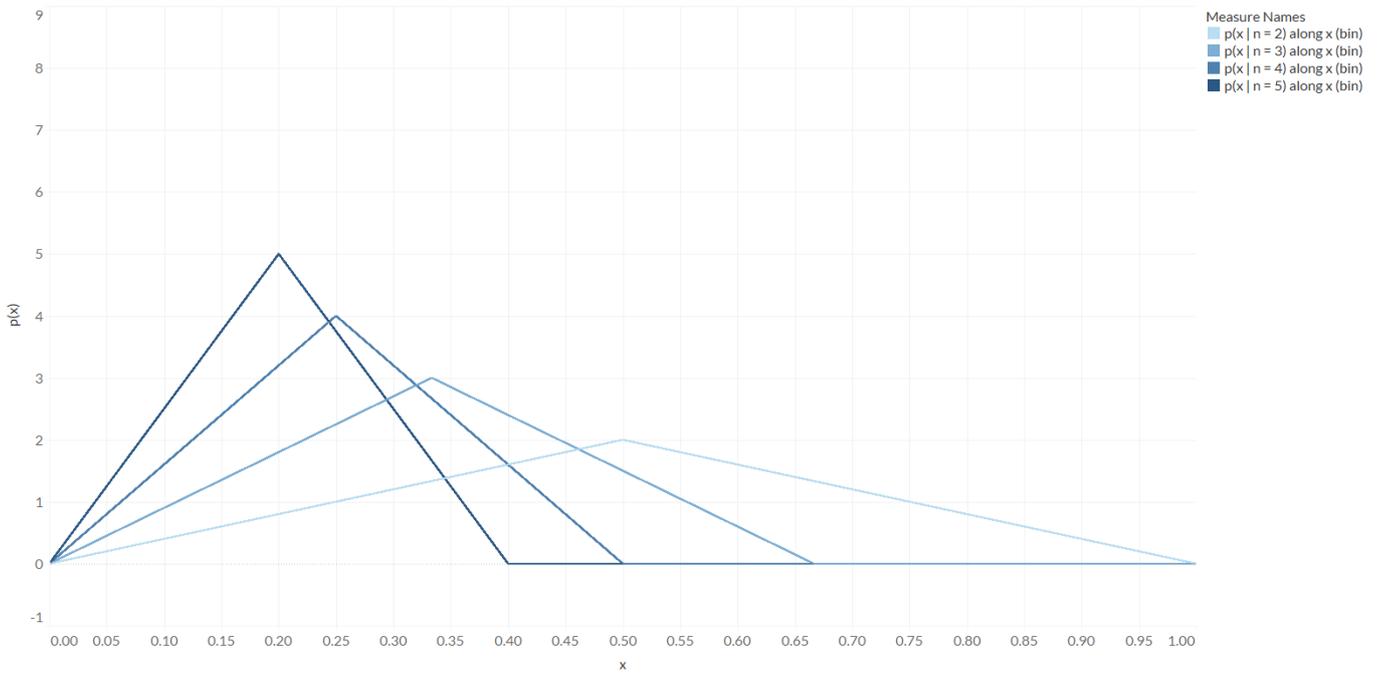

Figure 10: Balanced Shuffle spacing distributions for $n = [2, 3, 4, 5]$.

## 3.3 The Poláček Shuffle

While at Spotify in 2014, Lukáš Poláček wrote a shuffling algorithm, which we will refer to as the Poláček Shuffle, to address the inefficiencies of the Balanced Shuffle.[7] While the Balanced Shuffle effectively maps songs into intervals in $\mathbb{Z}$ and merges them together, the Poláček Shuffle maps songs into intervals in $\mathbb{R}$ instead without using spacers.

The Poláček Shuffle first performs the following steps on each song group:

1. Shuffle the song group with an unbiased shuffle.
2. Space the songs out evenly over $[0, 1]$ as if we're around a circle with a circumference of $1$.
3. Vary each song's position by sampling uniform r.v.s centered on $0$. The width of the uniform r.v.s is some fixed proportion of the distance between the evenly spaced songs.
4. Circular shift all positions around $[0, 1]$ with a uniform r.v. on $[0, 1]$.

Afterwards, the song groups are merged and sorted by song position.

This algorithm takes $O(n)$ space and $O(n\ log\ n)$, $O(n\ log\ k)$, or $O(n)$ time depending on if a comparison sort, $k$-way merge sort, or radix sort[10] is used, where $k$ is the number of song groups. This achieves a similar effect to the Balanced Shuffle while greatly reducing the complexity.

In essence, the Poláček Shuffle is identical to the Balanced Shuffle we extended from $\mathbb{Z}$ to $\mathbb{R}$ but with the ability to specify how far a song can vary from the center of its assigned interval and the addition of a circular shift. The spacing distribution of a song group is given by Equation 3, where $w \in [0, 1]$ specifies the width of the uniform r.v.s used to vary each song's position as a proportion of the distance between the evenly spaced songs, $\Pi(x)$ is the unit rectangle function, and $n$ is the number of songs in the group (Figures 11, 12).

$$p_{Poláček\ spacing}(x) = \frac{1}{n-1} \cdot 2 \left[ \frac{n}{w} \Pi \left( \frac{n}{w} x \right) * \left( -\frac{n}{w} \left( x - \frac{1}{n} \right) + \frac{1}{2} \right) \cdot \frac{n}{w} \Pi \left( \frac{n}{w} \left( x - \frac{1}{n} \right) \right) \right] + \frac{n-2}{n-1} \cdot \left[ \frac{n}{w} \Lambda \left( \frac{n}{w} \left( x - \frac{1}{n} \right) \right) \right] \tag{3}$$

The first term represents the spacing distributions between the items whose position distributions have been wrapped around from $1$ to $0$ and its adjacent unwrapped items while the second term represents the spacing distributions between consecutive unwrapped items.

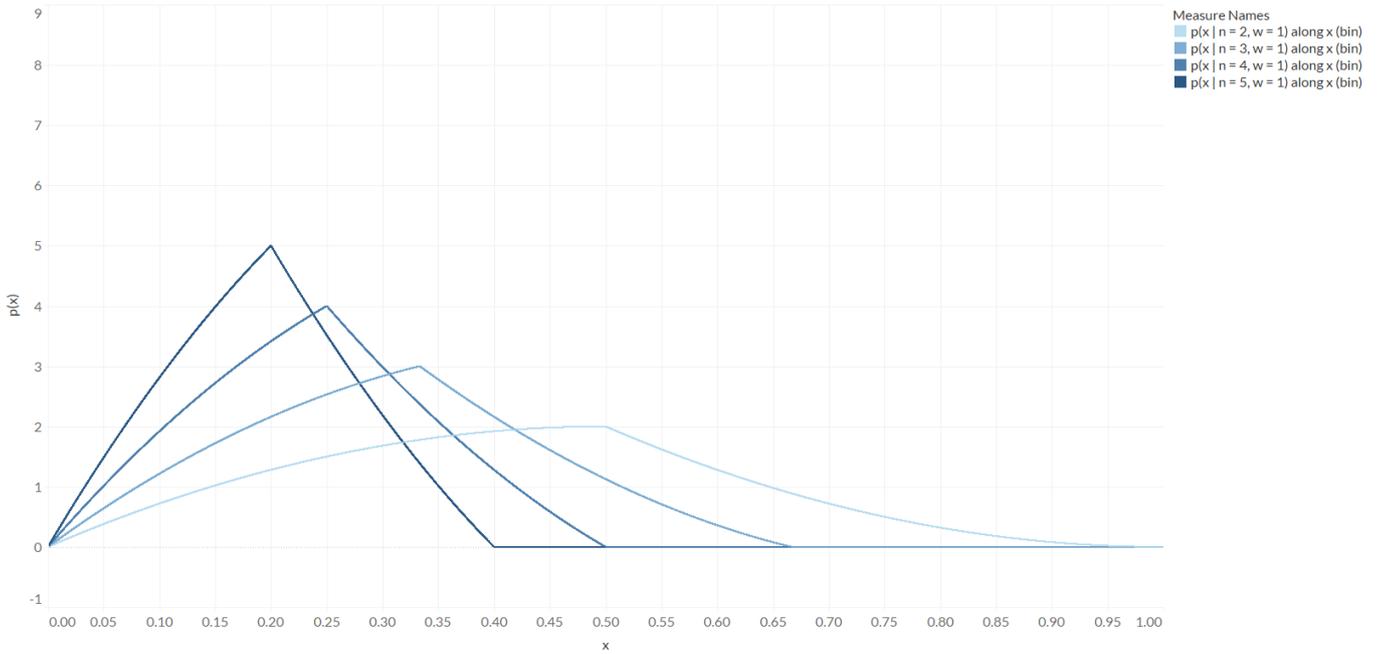

*Figure 11: Poláček Shuffle spacing distributions for $n = [2, 3, 4, 5]$ and $w = 1$.*

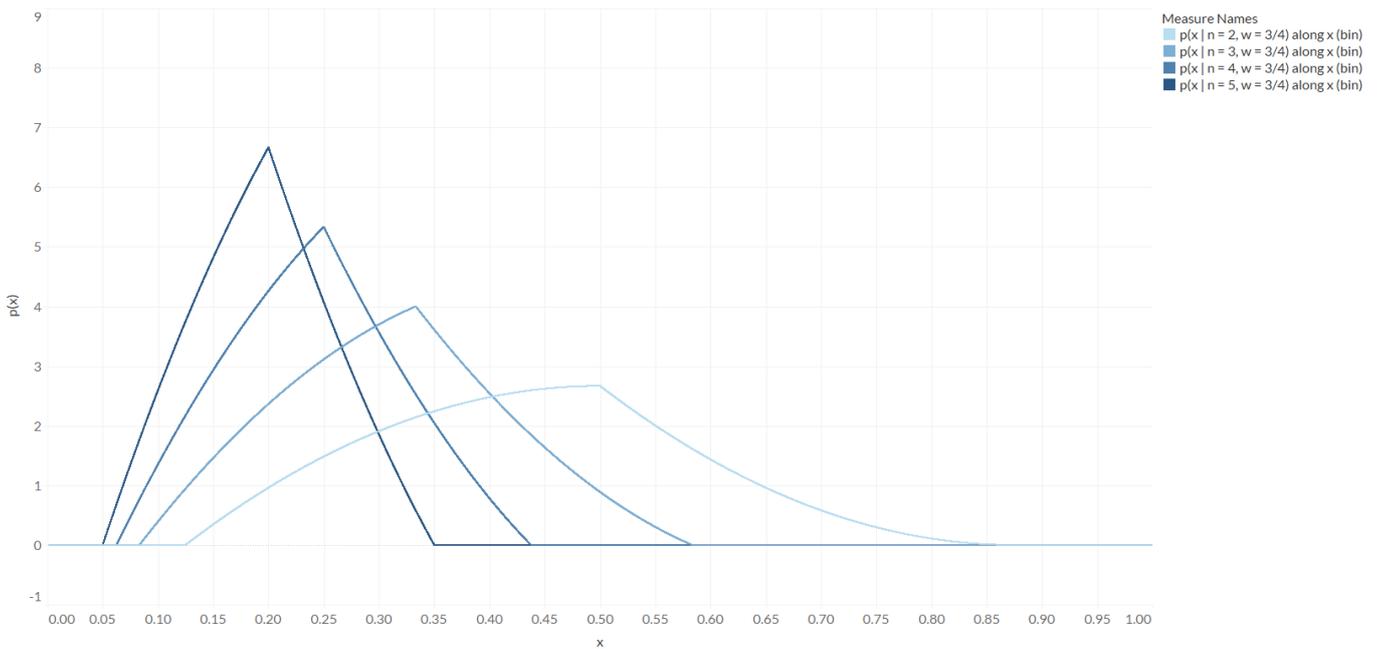

*Figure 12: Poláček Shuffle spacing distributions for $n = [2, 3, 4, 5]$ and $w = \frac{3}{4}$.*

The circular shift distorts the spacing distribution by potentially causing the position distribution of the last item to wrap around in front of the first. Shifting a middle item away from one neighboring item has the reciprocal effect of shifting it closer to its other neighbor, balancing out the average distance between items. Without the circular shift, the first and last items each only have one neighbor, so shifting away from their neighbor has no reciprocal effect. With the circular shift, however, the last item can wrap around and "change neighbors" if the r.v. used to vary its position pushes it too far from its original neighbor. This decreases the average distance between the boundary item and its sole neighbor.

# 4 Disordered Hyperuniform Systems

Looking at the spacing distributions for each of the reference shuffles, we see that unbiased shuffles make it most likely for songs in one group to cluster together while the Balanced Shuffle and Poláček Shuffle make it most likely for them to stay spread apart at some non-zero distance. Both of them do this by spacing the songs at even distances and then letting them shift around a little, essentially fixing item density at large length scales while letting it vary at smaller length scales.

This behavior is characteristic of disordered hyperuniform systems such as the distribution of cone cells in chicken eyes, the energy levels of heavy atomic nuclei, the eigenvalue distributions of various types of random matrices, the distribution of the nontrivial zeroes of the Riemann Zeta function, dynamical processes in ultra-cold atoms, receptor organization in the immune system, and many others which appear in a variety of biological, chemical, physical, and mathematical settings.[11] We'll take a brief look at the first three systems to get an idea of how disordered hyperuniform systems look in two dimensions and one dimension.

## 4.1 Chicken Eyes

One of the traits birds are best known for is their spectacular vision, demonstrated by some birds' ability to resolve tiny objects from exceptional distances. Optimal spatial sampling of light is generally expected from near-identical photoreceptors being arranged in a perfect lattice, such as the compound eyes of insects or the nearly crystalline eyes of some fish and reptiles.[12] Many birds such as chickens, however, use several differently-sized cone types which are often difficult, if not impossible, to arrange in a crystalline structure (Figure 13).

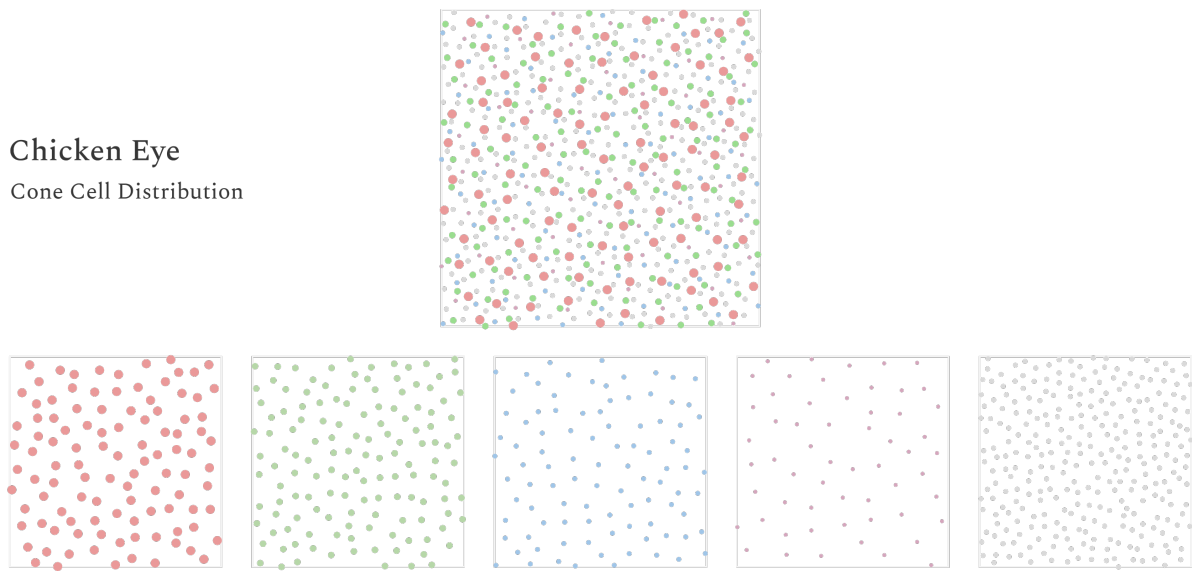

Figure 13: Distribution of cone cells (separated by type and combined) in a chicken eye.[12]

At first glance, there doesn't seem to be any apparent pattern in the photoreceptors. When considering each cone type separately, however, we see that cones of the same type are distributed such that they are neither too close nor too far apart. Each cone type is said to be distributed **hyperuniformly** with their combined distribution being **multi-hyperuniform**.[12]

## 4.2 Energy Levels of Heavy Atomic Nuclei

Light nuclei have simple structures and their individual energy levels can be calculated to extraordinary accuracy. Heavy nuclei, however, are far more complex and the calculation of their individual energy levels is impractical. Instead, it's more pragmatic to look at the collective behavior of the energy levels of heavy nuclei (Figure 14).

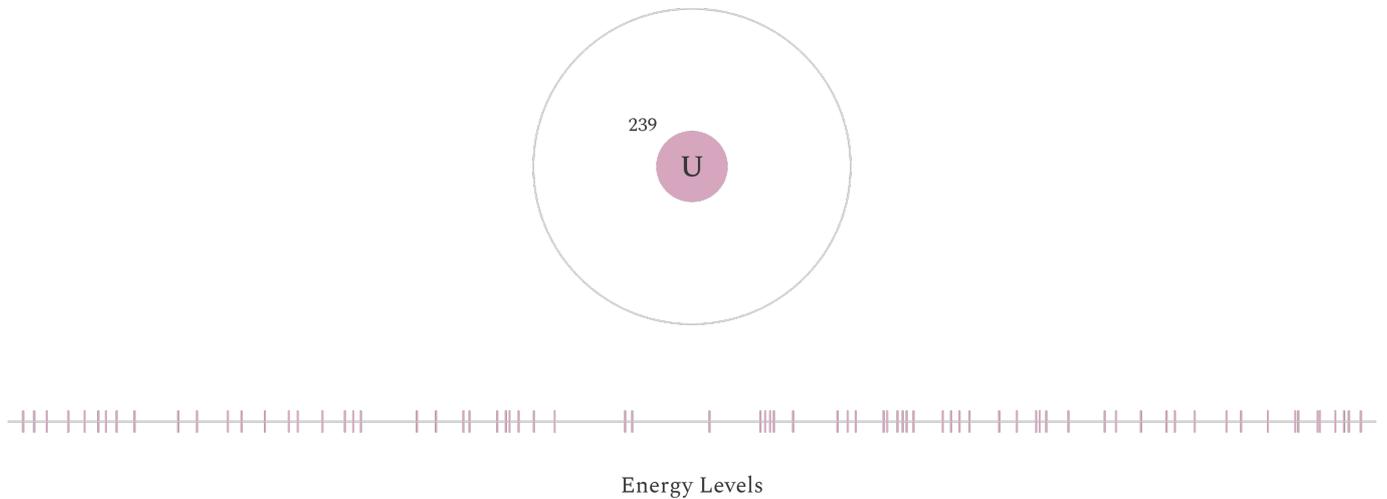

Figure 14: NNDC ENSDF adopted levels for Uranium-239 from $687.85\ keV$ to $1520.33\ keV$.[13]

Nobel laureate Eugene Wigner suggested using random matrices to model this behavior.[14] Special types of random matrices have eigenvalues, a set of characteristic numbers, which are distributed hyperuniformly on the number line.

## 4.3 Gaussian Ensembles

In his 1967 paper, Wigner proposed using Gaussian ensembles to model the energy level distributions of heavy atomic nuclei.[14] The Gaussian ensembles are families of matrices whose elements are all independent and identically distributed (i.i.d.) Gaussian random variables (i.e. normally distributed).

Separate names are given to families with certain matrix symmetries such as the Gaussian Orthogonal Ensemble (GOE) for real symmetric matrices, the Gaussian Unitary Ensemble (GUE) for complex Hermitian matrices, and the Gaussian Symplectic Ensemble (GSE) for quaternionic Hermitian matrices.[15]

The eigenvalue spacing distributions for these matrix ensembles are approximated by the Wigner Surmise which is defined on $[0, \infty)$. Figure 15 shows the Wigner Surmise for Dyson indices $\beta = [1, 2, 4]$, where the GOE corresponds to $\beta = 1$, the GUE to $\beta = 2$, and the GSE to $\beta = 4$.

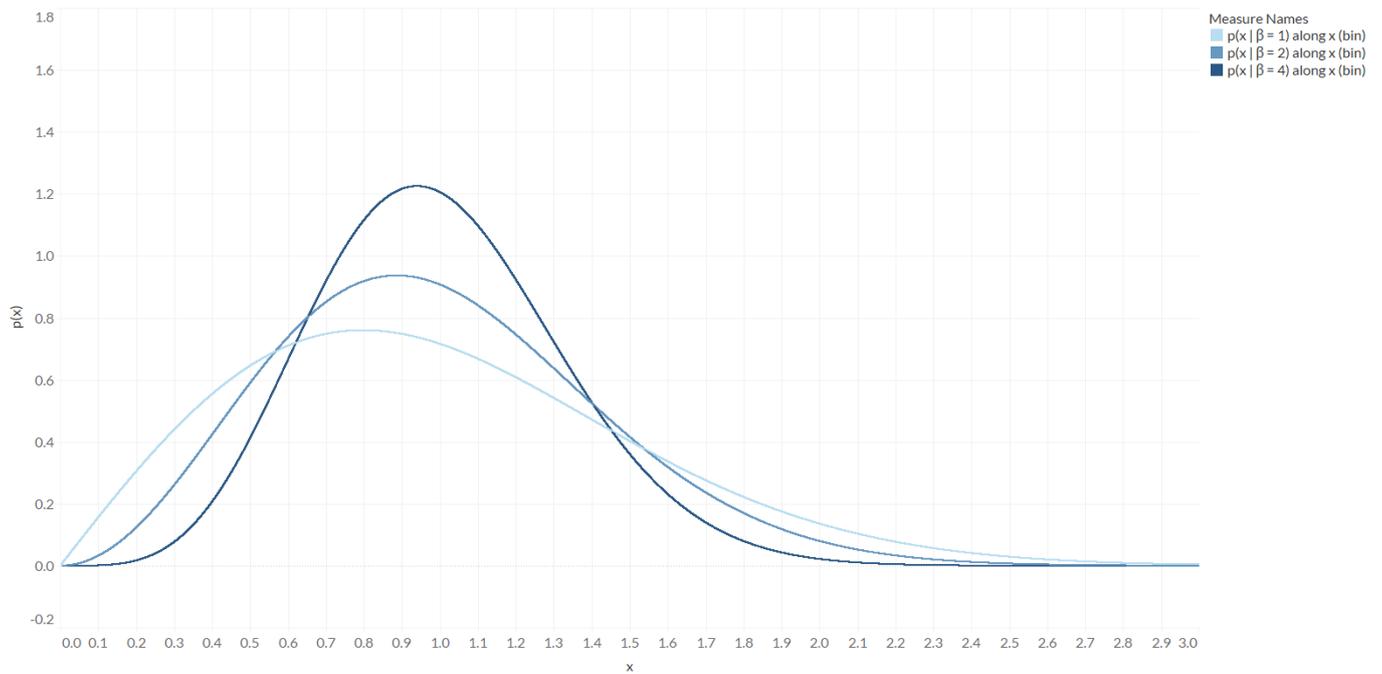

*Figure 15: Wigner Surmise for $\beta = [1, 2, 4]$.*

Much like the spacing distributions for the Balanced Shuffle and Poláček Shuffle, the eigenvalues of these matrices are most likely to stay spread apart at some non-zero distance. One interesting note is that the Wigner Surmise looks more and more Gaussian as $\beta$ increases from $1$ to $4$. Since the distance between two Gaussian r.v.s is another Gaussian r.v.,[16] this hints that each of the $i$-th eigenvalues may be Gaussian-like r.v.s. We can test this by doing spectral decomposition on matrices of fixed dimensions from the GUE (Figures 16, 17, 18, 19, 20, 21). We won't be using the GSE because most linear algebra libraries only support the spectral decomposition of real and complex matrices but not quaternionic matrices.

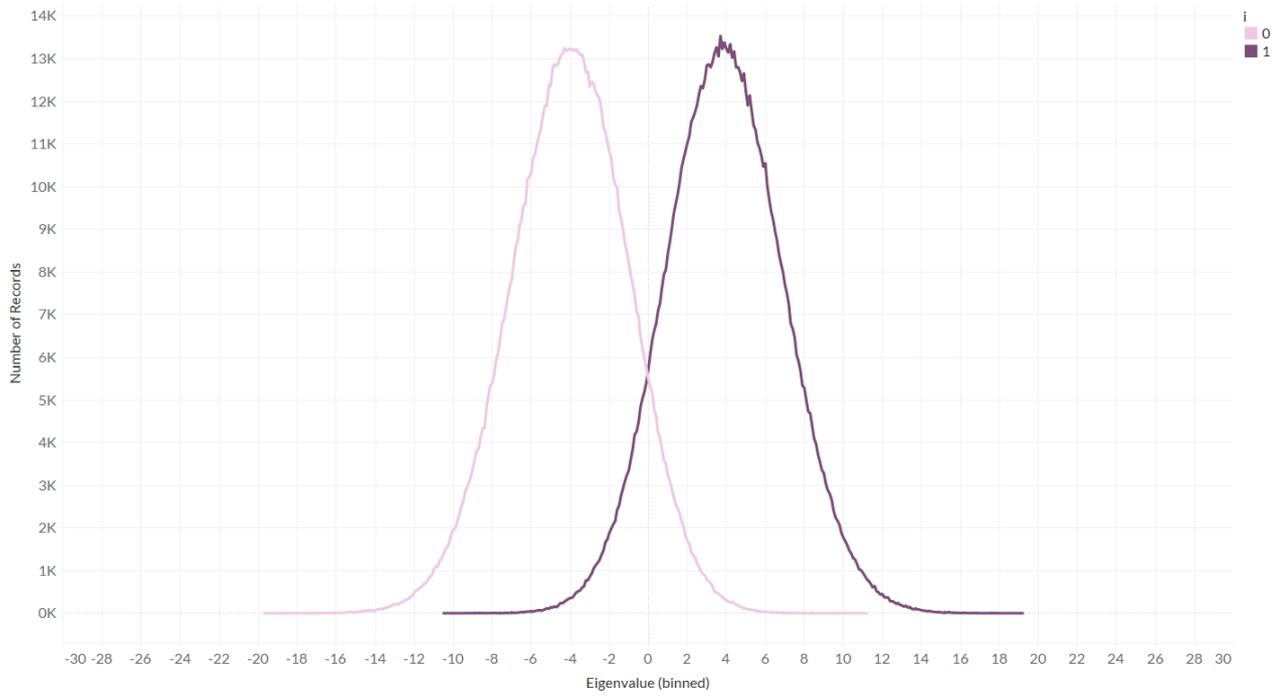

Figure 16: $i$-th eigenvalue distributions for $10^6$ samples of $2 \times 2$ matrices from the GUE coarse-scaled[17] to $[-10, 10]$.

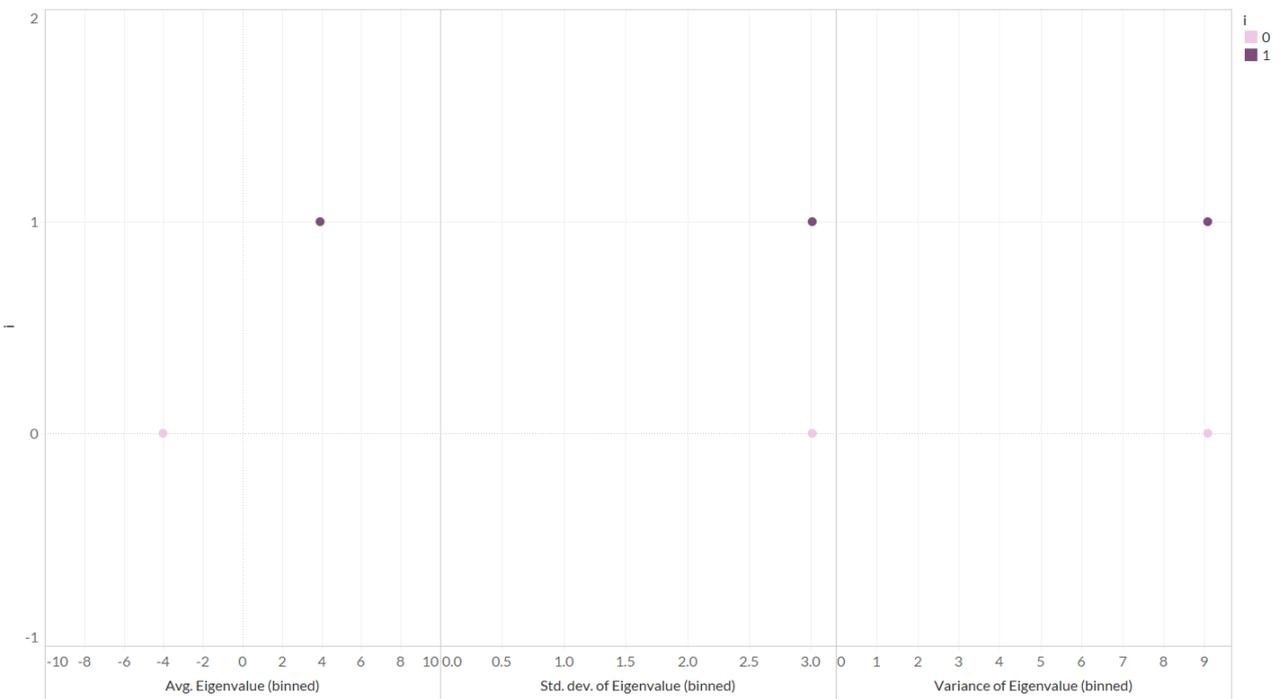

Figure 17: $i$-th eigenvalue statistics for $10^6$ samples of $2 \times 2$ matrices from the GUE coarse-scaled[17] to $[-10, 10]$.

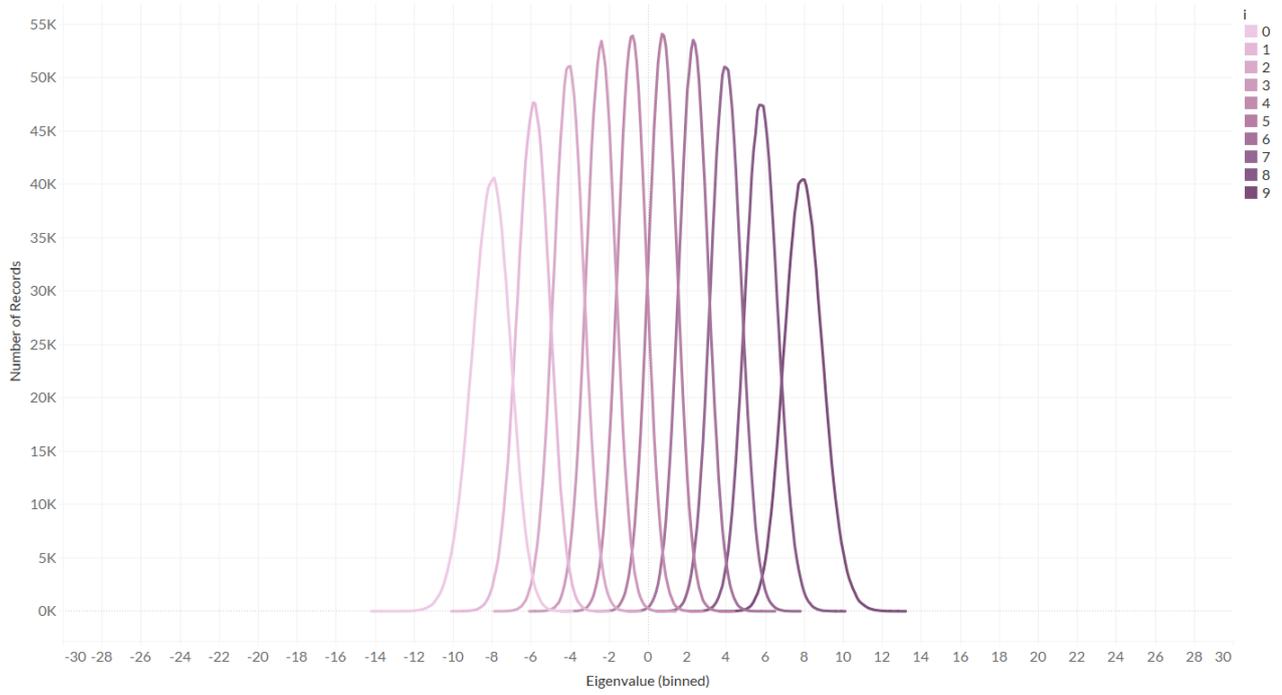

Figure 18: $i$-th eigenvalue distributions for $10^6$ samples of $10 \times 10$ matrices from the GUE coarse-scaled[17] to $[-10, 10]$.

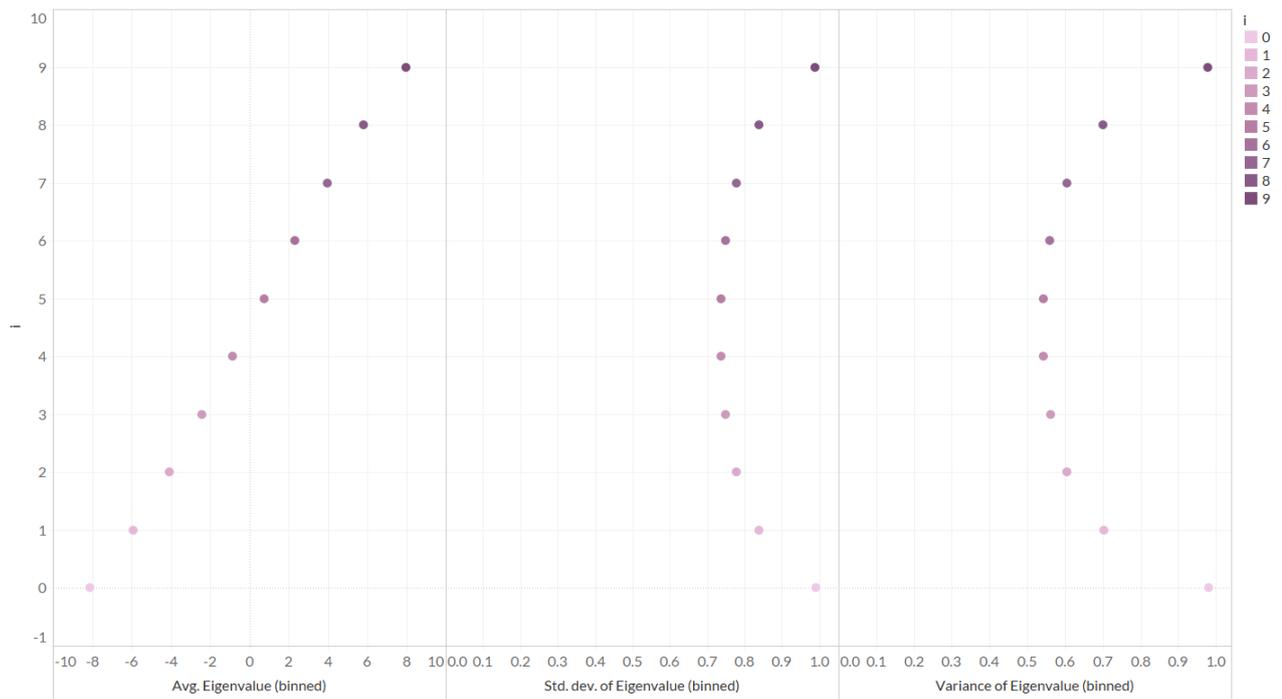

Figure 19: $i$-th eigenvalue statistics for $10^6$ samples of $10 \times 10$ matrices from the GUE coarse-scaled[17] to $[-10, 10]$.

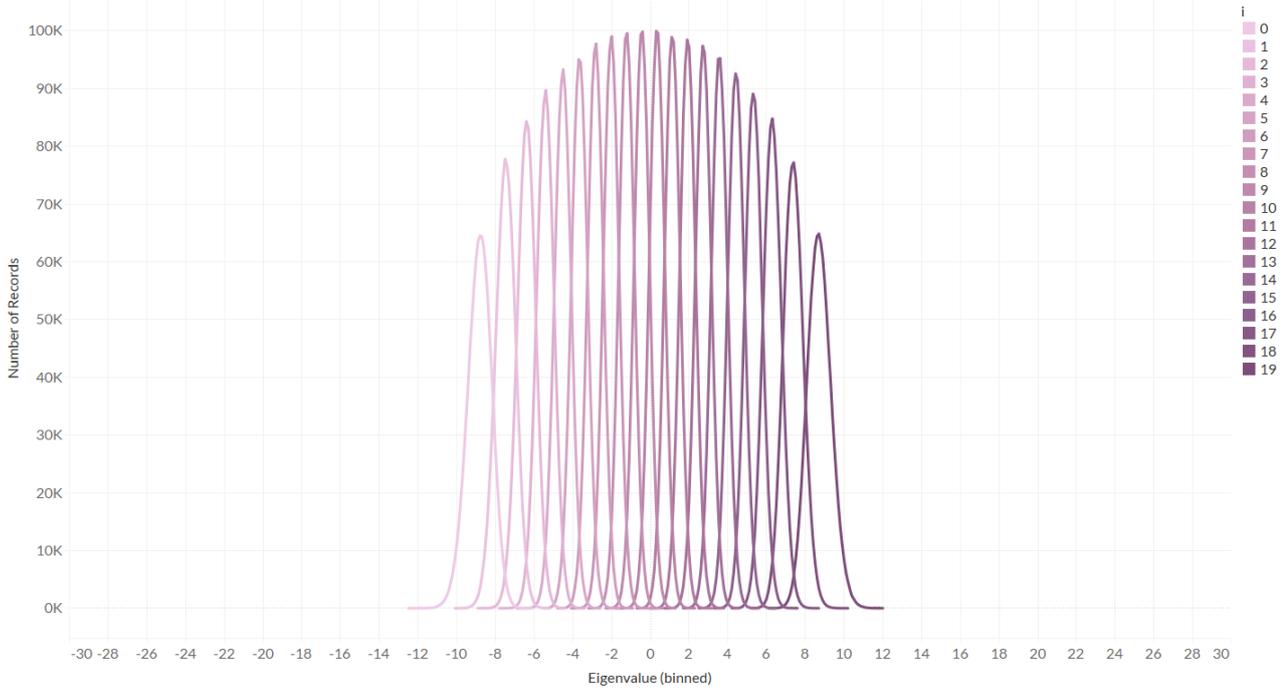

*Figure 20: $i$-th eigenvalue distributions for $10^6$ samples of $20 \times 20$ matrices from the GUE coarse-scaled[17] to $[-10, 10]$.*

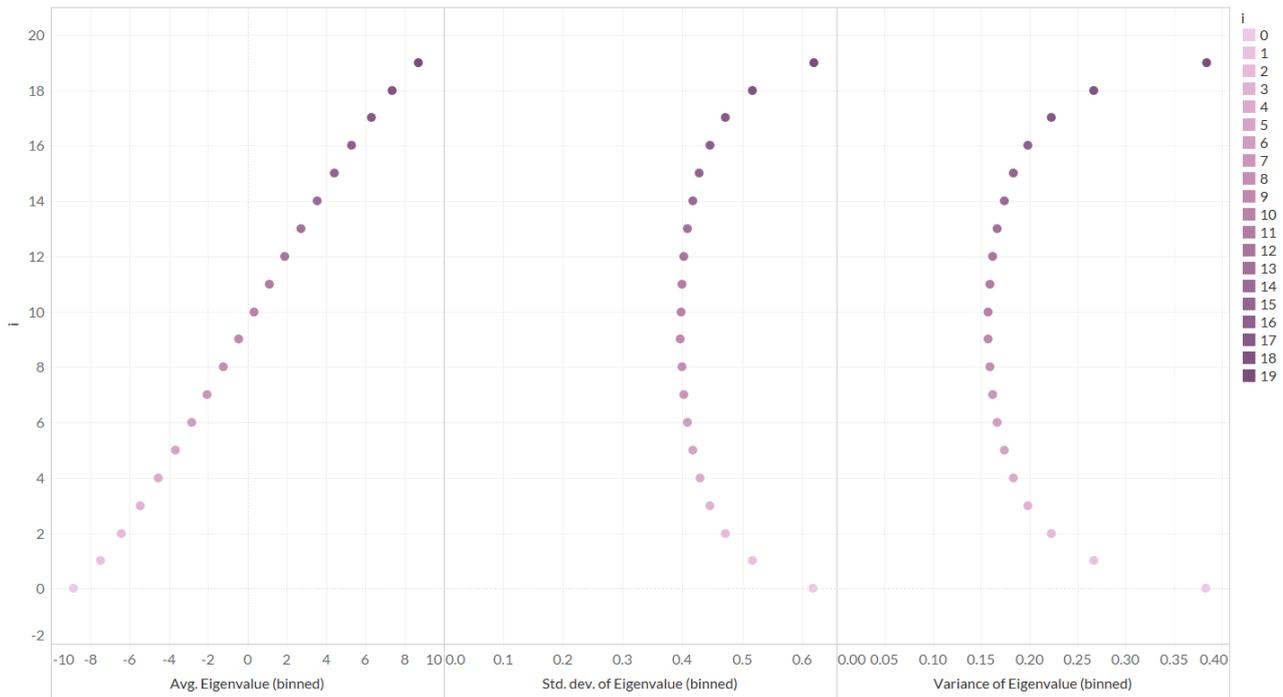

*Figure 21: $i$-th eigenvalue statistics for $10^6$ samples of $20 \times 20$ matrices from the GUE coarse-scaled[17] to $[-10, 10]$.*

To our pleasant surprise, they are Gaussian-like r.v.s but with increasing inter-peak distances and variances the further they are from the center as if they are more "loosely bound" the further they are from the center. This will be useful later for creating more efficient approximations of a Gaussian ensemble based shuffle.

# 5 Cluster Diffusing (CD) Shuffles

Looking at how the reference shuffles are structured, we can create a general set of steps for our shuffle to follow:

1. **Alter**
   Reorder the items in each item group.

2. **Map**
   Assign each item to a number.

3. **Merge**
   Combine the item groups and sort the items by their assigned number.

We'll be looking at these components in reverse order since the merge and map steps are essential to shuffle play while the alter step is only particularly meaningful for shuffle + repeat play.

## 5.1 Merge

The merge step combines the item groups and sorts the combined items by their assigned number. We can do this in one of three ways:

1. Concatenating the item groups and then sorting the items with a comparison sort.

2. Combining the presorted item groups with a $k$-way merge sort.

3. Concatenating the item groups and then sorting the items with a radix sort.

The first approach takes $O(n)$ time to concatenate the item groups and $O(n \, log \, n)$ time to sort them with a comparison sort, giving an overall time complexity of $O(n \, log \, n)$.

The second approach takes $O(n \, log \, k)$ time to merge sort the groups but requires each group to already be sorted.

The third approach is the fastest method, taking $O(n)$ time both to concatenate the groups and to sort the combined items, giving an overall time complexity of $O(n)$.

The space requirement for all of these approaches is $O(n)$ because we'll be copying each item (or its reference) from the input to the output. This is done to preserve information about the item ordering of each group which is used for shuffle + repeat play, something we will discuss in the alter step.

## 5.2 Map

To mimic multi-hyperuniform systems and to reduce the probability of same group item clusters in the resulting sequence, our shuffle should aim to space items from the same group at relatively even distances across the output sequence. Some amount of variability should be present in the item positions to prevent each group's items from occupying the same positions across all shuffles and creating a lattice-like appearance as a result.

Most of the mapping techniques described below do this by evenly spacing out points over some interval and then varying them slightly with r.v.s. What these maps do to the shape of the overall item position distribution is akin to creating a crenellated wall by adding merlons, which are the vertical protrusions found on the top of rooks (Figure 22) or the walls of many kinds of ancient defensive architecture (e.g. the Great Wall of China).

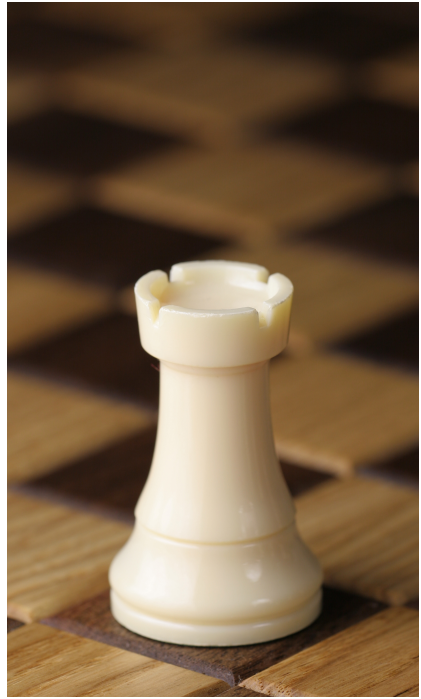

*Figure 22: A rook from chess with a crenellated top.[18]*

For a small visual demonstration of each map, example outputs of a CD shuffle using each map on the input in Figure 23 will be shown.

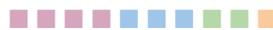

*Figure 23: The example input used for the example outputs of a CD shuffle using each map.*

### 5.2.1 Lattice Map

The most basic map is the lattice map, which simply spreads each group's items out evenly over some interval without varying their positions like a Balanced Shuffle that doesn't vary the item positions. We can find the positions using Equation 4, where $i$ is the zero-based index of the $i$-th item, $n$ is the number of songs in the group, and $r$ is the radius of the interval the items are spread over (i.e. $[-r, r]$).

$$p_{position}(i) = \frac{r}{n} \cdot (1 - n + 2i) \qquad (4)$$

Assuming that items from different groups mapped to the same number are sorted in some consistent, stable manner, this map produces the same output sequence for all shuffles of the same set of input groups.

Overall, this map takes $O(n)$ time and $O(n)$ space.

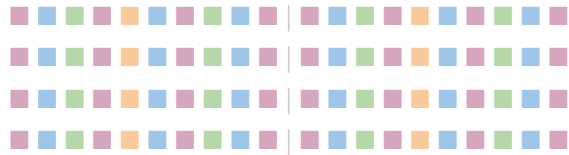

*Figure 24: A set of example outputs from a CD shuffle using the lattice map.*

### 5.2.2 Spectral Map

The spectral map uses the eigenvalues of matrices from the GUE to assign item positions. For each group, we create a random $n \times n$ matrix from the GUE, where $n$ is the number of items in the group. This matrix is coarse-scaled so that its eigenvalues lie approximately within the same interval centered on $0$ for all groups (e.g. $[-1, 1]$). Afterwards, we perform spectral decomposition on the matrix and assign the $i$-th item to the $i$-th eigenvalue.

Spectral decomposition a Hermitian matrix, and thus this map, takes $O(n^3)$ time and $O(n^2)$ space.

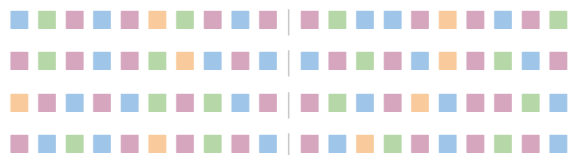

*Figure 25: A set of example outputs from a CD shuffle using the spectral map.*

### 5.2.3 Gaussian Map

The high time and space complexities of the spectral map make it very impractical for everyday use. Fortunately, the Gaussian-like nature of the individual eigenvalues gives us some ideas for how to create a more efficient approximation.

To avoid the complexities of trying to mimic the changing inter-peak distances and variances per item, we'll spread the items out evenly over some interval centered on $0$ and then vary their positions using Gaussian r.v.s with the same variance. This is akin to the Balanced Shuffle extended from $\mathbb{Z}$ to $\mathbb{R}$ but with Gaussian r.v.s instead of uniform r.v.s.

Since the standard deviation $\sigma$ of the Gaussian-like eigenvalues decreases as the number of items in the group increases, we'll specify $\sigma$ as some proportion of the width between the evenly spaced items with the parameter $w$. This is similar to how the Poláček Shuffle specifies the width of the uniform r.v.s used to vary song position as a proportion of the distance between the evenly spaced songs with its own $w$ parameter. We'll use a $w$ of $\frac{1}{2}$ so that each item's varied position will be within the interval whose endpoints are the midpoints between itself and its two neighbors' initial evenly spaced positions $\sim 95\%$ of the time (the $68 - 95 - 99.7$/empirical/three-sigma rule).

With all of these combined, we get that the position distribution for item $i$ in a zero-indexed group with $n$ items is given by Equation 5, where $r$ is the approximate radius the items are spread over (i.e. $[-r, r]$).

$$p_{position}(i) = \frac{r}{n} \cdot (1 - n + 2i + X_{Gaussian}(\mu, \sigma)) \qquad \mu = 0, \sigma = \frac{1}{2} \qquad (5)$$

Since the Gaussian distribution has an infinite support, we need to sort the positions before assigning them because one Gaussian r.v. with a greater mean than another isn't guaranteed to produce a greater valued sample than the other.

Thus, this map takes $O(n \log n)$ or $O(n)$ time (depending on if a comparison sort or radix sort is used) and $O(n)$ space.

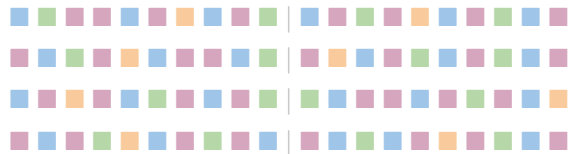

*Figure 26: A set of example outputs from a CD shuffle using the Gaussian map.*

### 5.2.4 Von Mises Map

One potential improvement we could make from the Gaussian map is varying the item positions with non-overlapping r.v.s instead of the overlapping ones in the Gaussian map. This removes the need to sort the positions before assigning them to the items while also preventing some large, low probability clusters from forming. Conveniently, there exists the notion of a wrapped normal distribution which has a finite support and is the result of "wrapping" the normal distribution around the unit circle. This distribution is closely approximated by the von Mises distribution which is often used instead due to being simpler to work with.[19]

The von Mises distribution is defined on $[-\pi, \pi]$ and, instead of a standard deviation parameter $\sigma$, has a concentration parameter $\kappa$ which is analogous to $\frac{1}{\sigma^2}$. To rescale the distribution to be on $[-1, 1]$ while trying to balance out the resulting variance change, we'll divide the von Mises r.v. by $\pi$ and divide $\kappa$ by $\pi^2$. With this, we get that the position distribution for the $i$-th item in a zero-indexed group with $n$ items is given by Equation 6, where $r$ is the radius the items are spread over (i.e. $[-r, r]$).

$$p_{position}(i) = \frac{r}{n} \cdot \left(1 - n + 2i + \frac{X_{von\ Mises}(\mu, \kappa)}{\pi}\right) \qquad \mu = 0, \kappa = \left(\frac{2}{\pi}\right)^2 \qquad (6)$$

Overall, this map takes $O(n)$ time and $O(n)$ space.

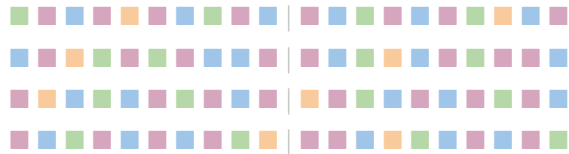

Figure 27: A set of example outputs from a CD shuffle using the von Mises map.

### 5.2.5 Benchmarks

Playlist compositions can widely vary between not only different playlists but also within a single playlist depending on how it's divided into groups. To see how these different maps perform, we'll use a set of benchmark playlists with different sizes and group size distributions (Table 2).

| Group Size *(Count)* | | Group Size Distribution | | | |
|---|---|---|---|---|---|
| | | Impulse | Uniform | Binomial | Zipf |
| **Playlist Size** | Tiny | 2 *(5)* | 1, 2, 3, 4 | 1, 2 *(2)*, 3 | 6, 3, 2 |
| | Small | 3 *(7)* | 1, 2, ..., 6 | 1, 2 *(3)*, 3 *(3)*, 4 | 12, 6, 4, 3 |
| | Medium | 5 *(11)* | 1, 2, ..., 10 | 1, 2 *(4)*, 3 *(6)*, 4 *(4)*, 5 | 24, 12, 8, 6 |
| | Large | 8 *(17)* | 1, 2, ..., 16 | 1, 2 *(5)*, 3 *(10)*, 4 *(10)*, 5 *(5)*, 6 | 60, 30, 20, 15, 12 |

*Table 2: The playlist sizes and group size distributions of the benchmark playlists.*

The impulse group size distribution divides the playlist into equally sized groups and is named "impulse" because it appears as an impulse on a group size v. count graph.

The uniform group size distribution has one group of each size up to a certain size and is named as is because it looks like a uniform distribution on a group size v. count graph.

The binomial group size distribution has group sizes distributed binomially with $p = 0.5$. This approximates a playlist where group sizes are distributed in a somewhat normal fashion.

The Zipf group size distribution splits songs among groups such that when groups are ranked by decreasing size, the number of songs in a group relative to the largest group is proportional to $1$ over its rank. This is in reference to Zipf's law and the prevalence of power laws in a variety of settings.

We'll be looking at the cluster size distributions, average cluster sizes, and cluster location distributions of an unbiased shuffle map, the Balanced Shuffle's map, the Poláček Shuffle's map, and CD shuffles using each of the different maps for **two consecutive shuffles** (Figures 28, 29, 30, 31, 32, 33, 34, 35, 36, 37, 38, 39). Using two consecutive shuffles lets us see if a map tends to create clusters at where shuffles meet, something we would like to avoid because it exacerbates the transition between shuffles in shuffle + repeat play.

## Cluster Sizes

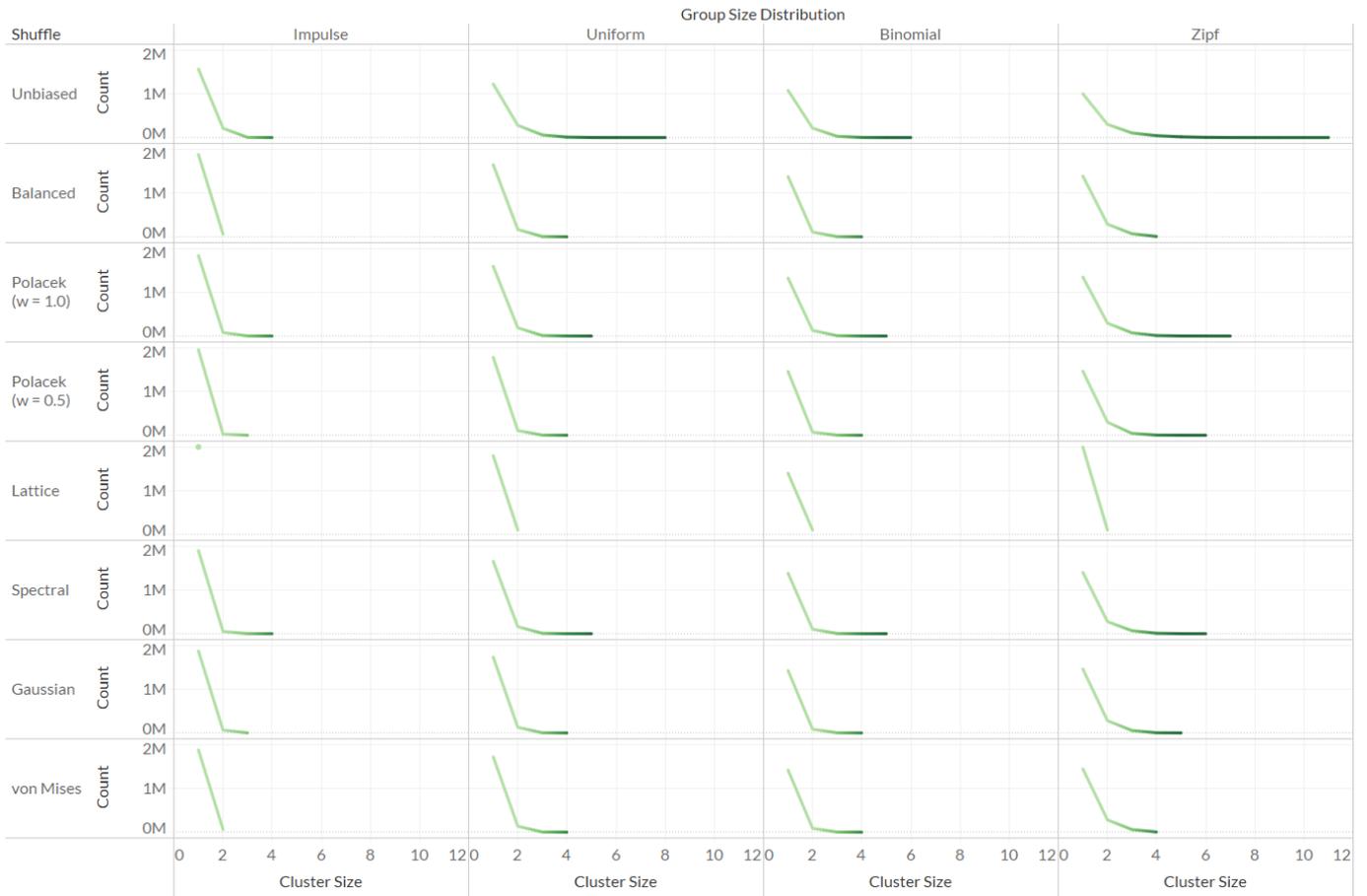

*Figure 28: Cluster size distributions for $10^5$ consecutive shuffle pairs on each of the **tiny** benchmark playlists using each of the maps.*

## Average Cluster Sizes

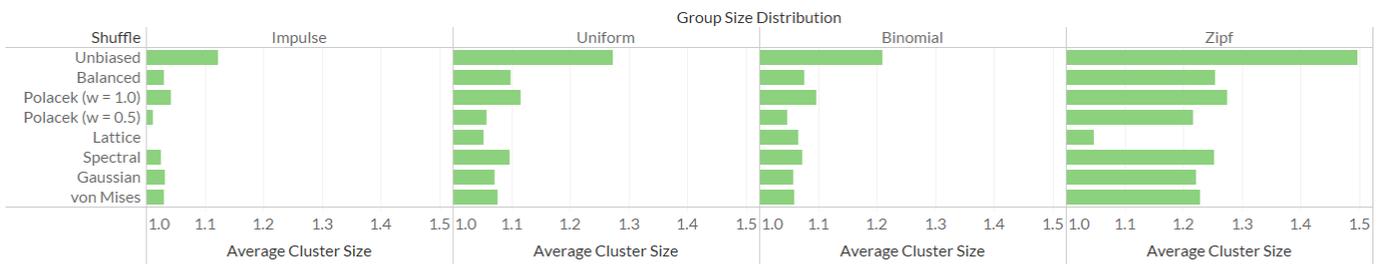

*Figure 29: Average cluster sizes for $10^5$ consecutive shuffle pairs on each of the **tiny** benchmark playlists using each of the maps.*

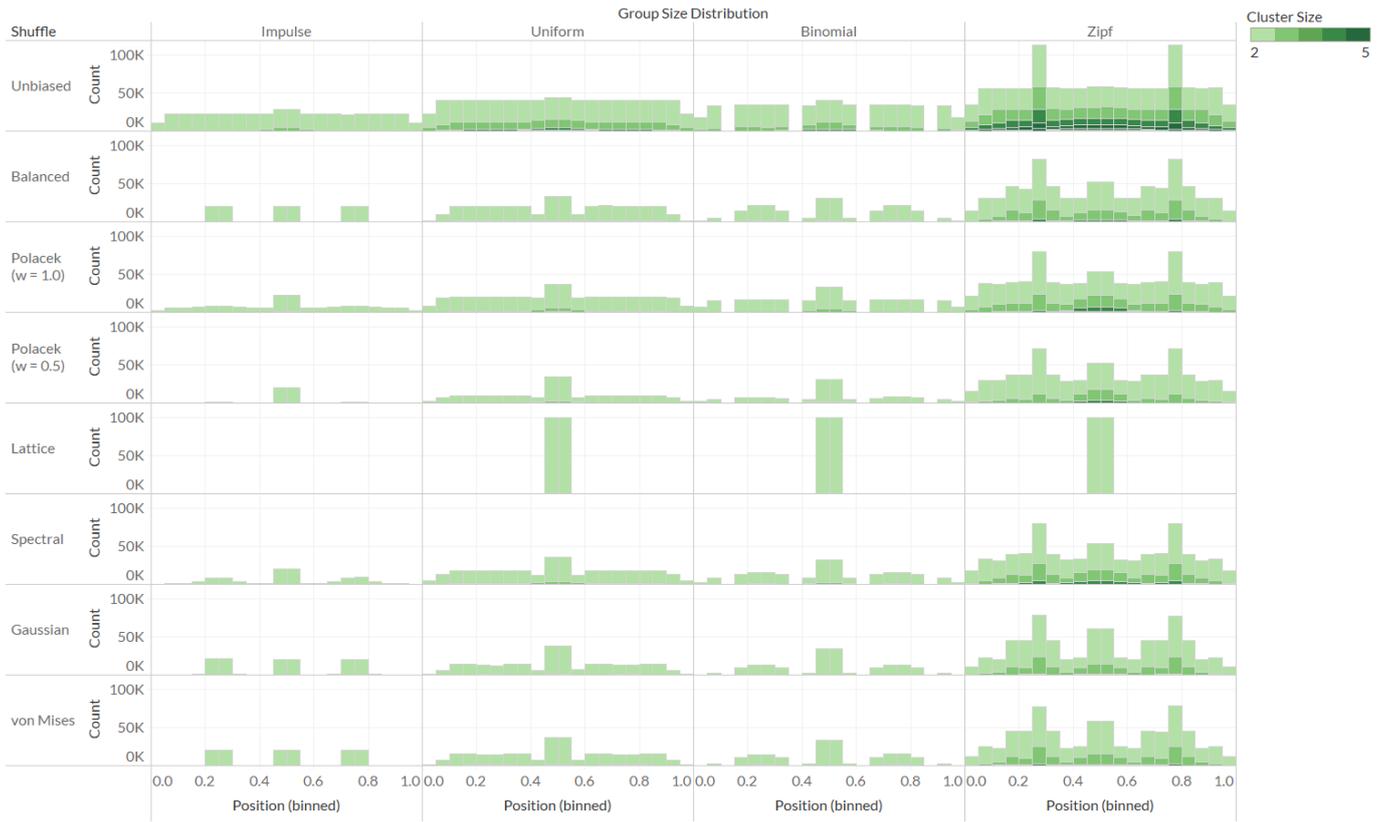

*Figure 30: Cluster location distributions for $10^5$ consecutive shuffle pairs on each of the **tiny** benchmark playlists using each of the maps. Bins are $0.05$ wide.*

Cluster Sizes

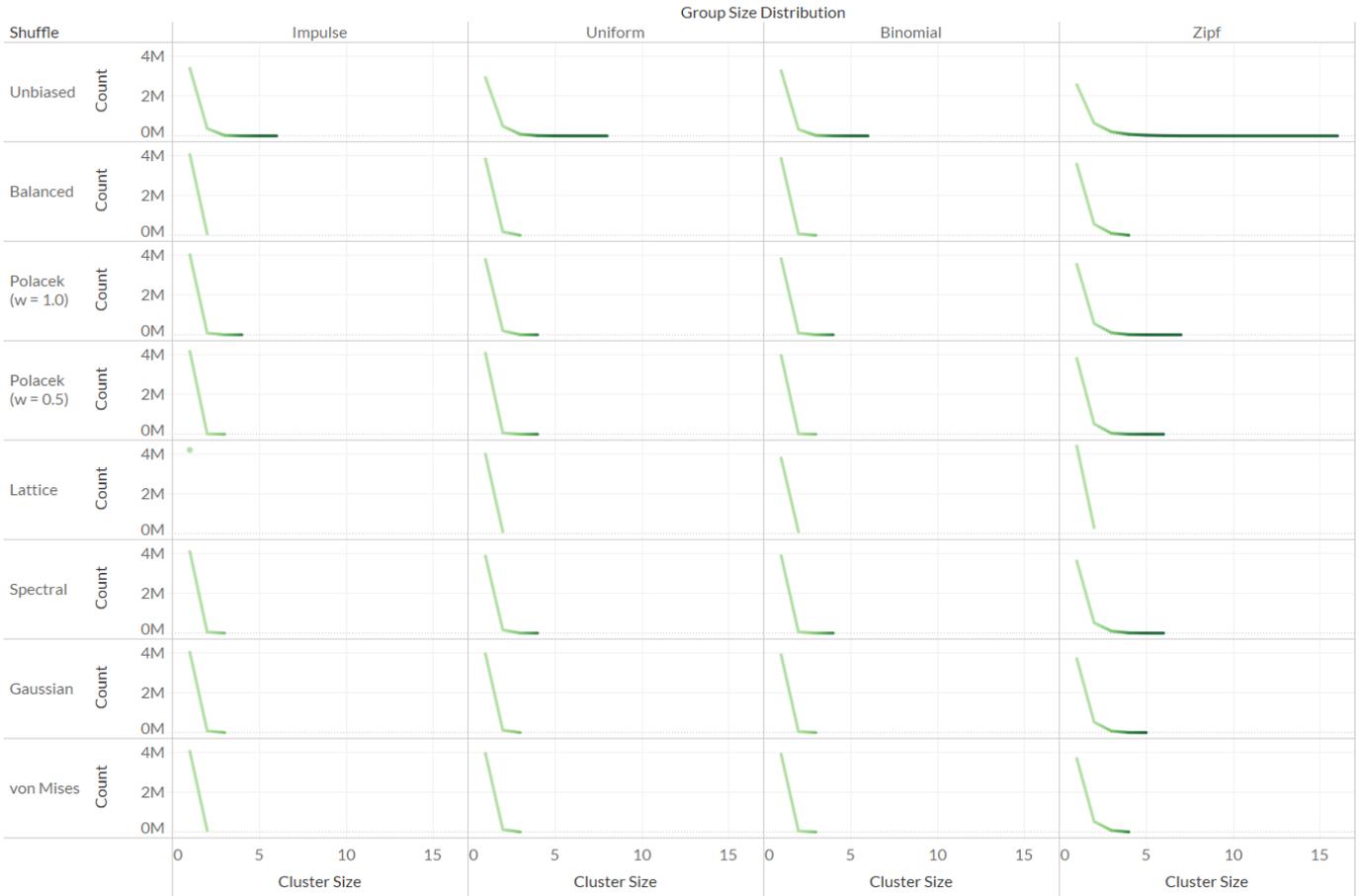

Figure 31: Cluster size distributions for $10^5$ consecutive shuffle pairs on each of the **small** benchmark playlists using each of the maps.

Average Cluster Sizes

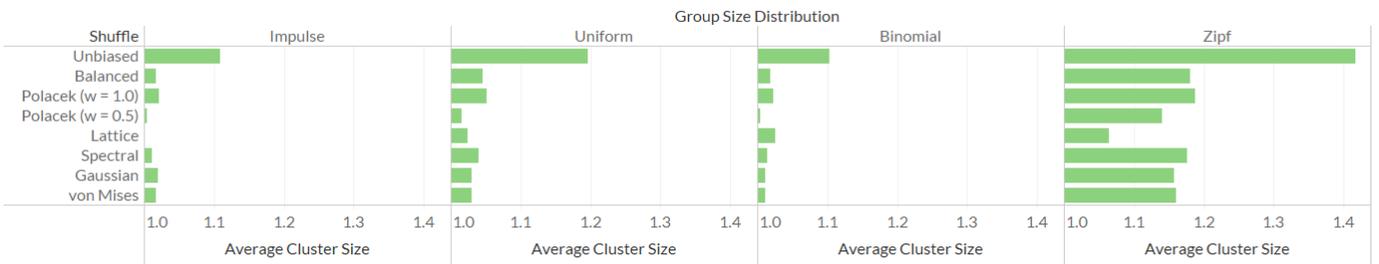

Figure 32: Average cluster sizes for $10^5$ consecutive shuffle pairs on each of the **small** benchmark playlists using each of the maps.

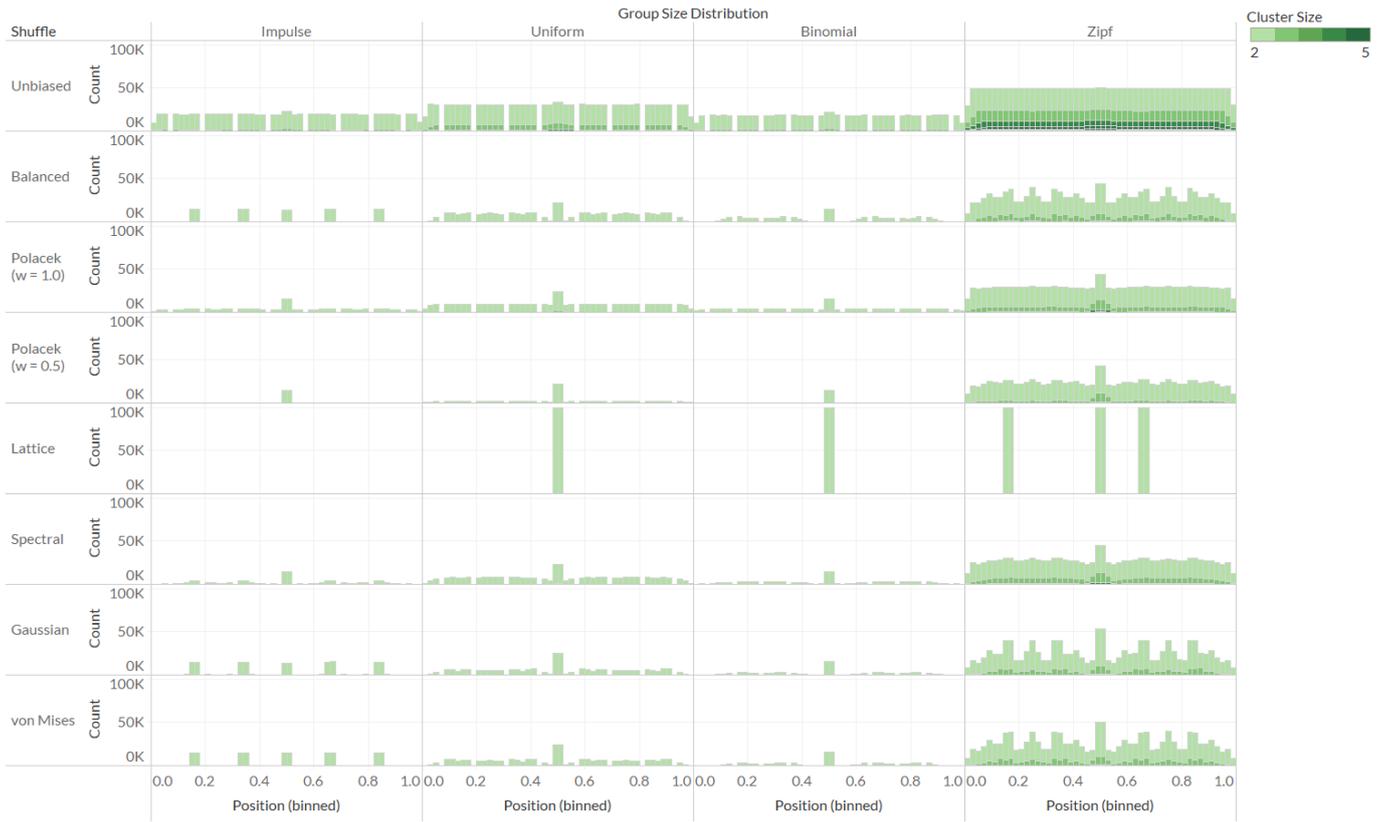

Figure 33: Cluster location distributions for $10^5$ consecutive shuffle pairs on each of the **small** benchmark playlists using each of the maps. Bins are $0.02$ wide.

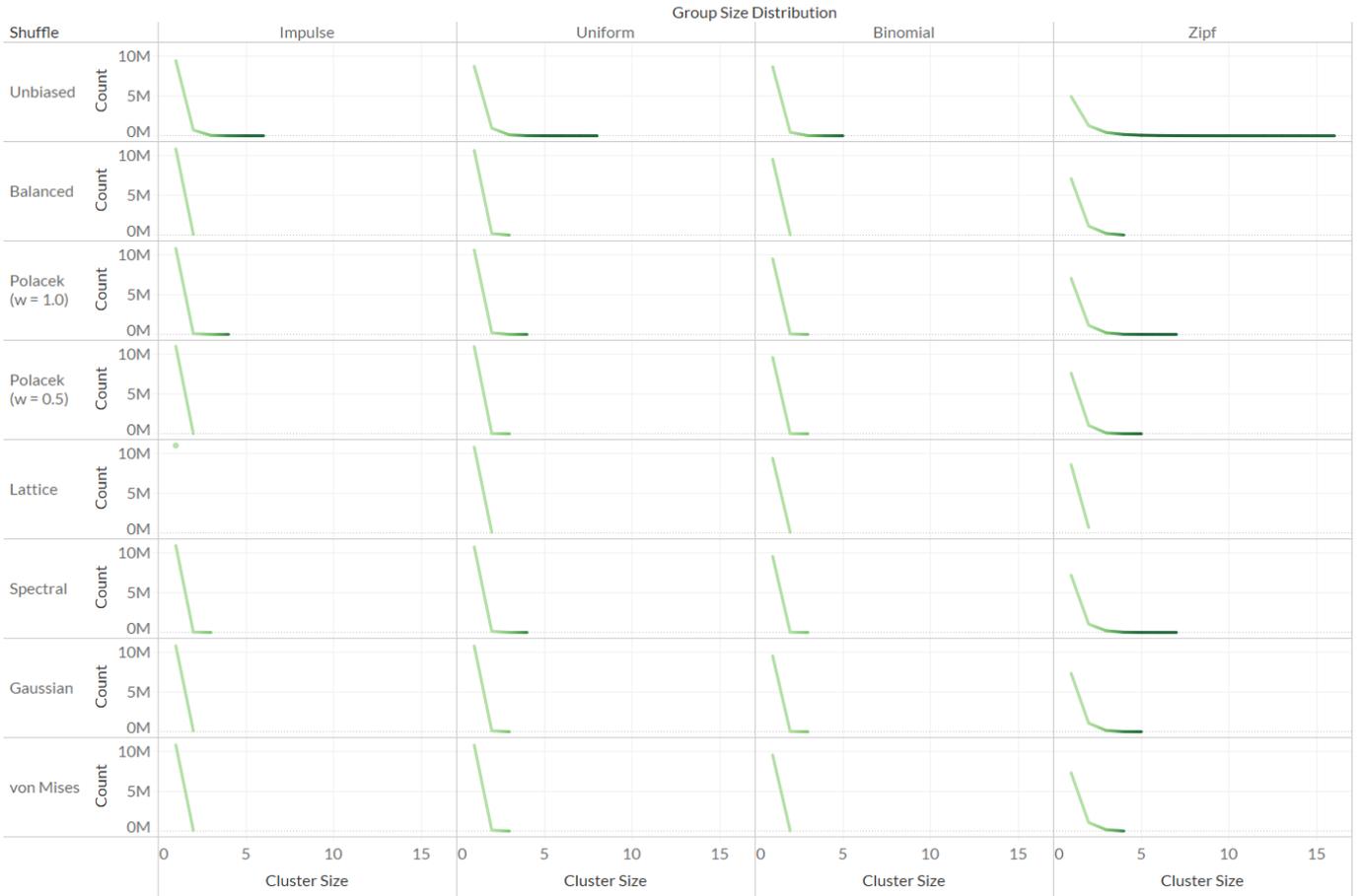

*Figure 34: Cluster size distributions for $10^5$ consecutive shuffle pairs on each of the **medium** benchmark playlists using each of the maps.*

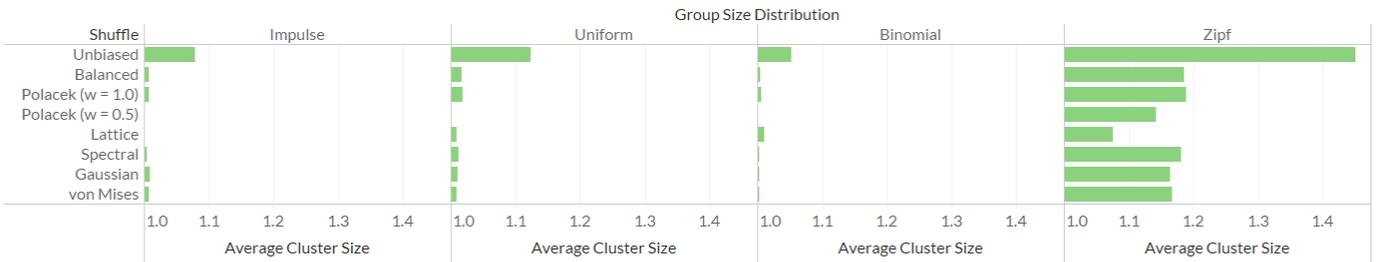

*Figure 35: Average cluster sizes for $10^5$ consecutive shuffle pairs on each of the **medium** benchmark playlists using each of the maps.*

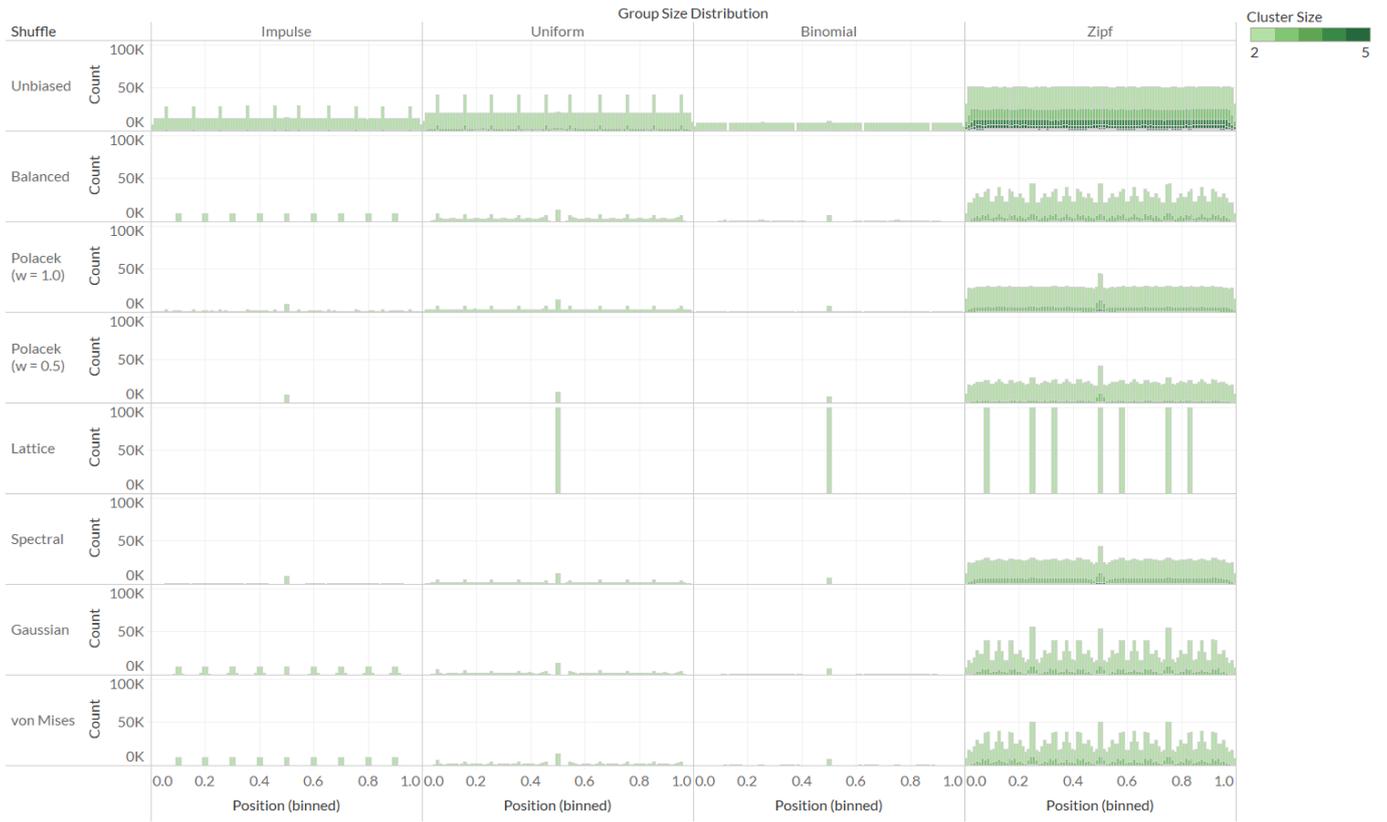

*Figure 36: Cluster location distributions for $10^5$ consecutive shuffle pairs on each of the **medium** benchmark playlists using each of the maps. Bins are $0.01$ wide.*

## Cluster Sizes

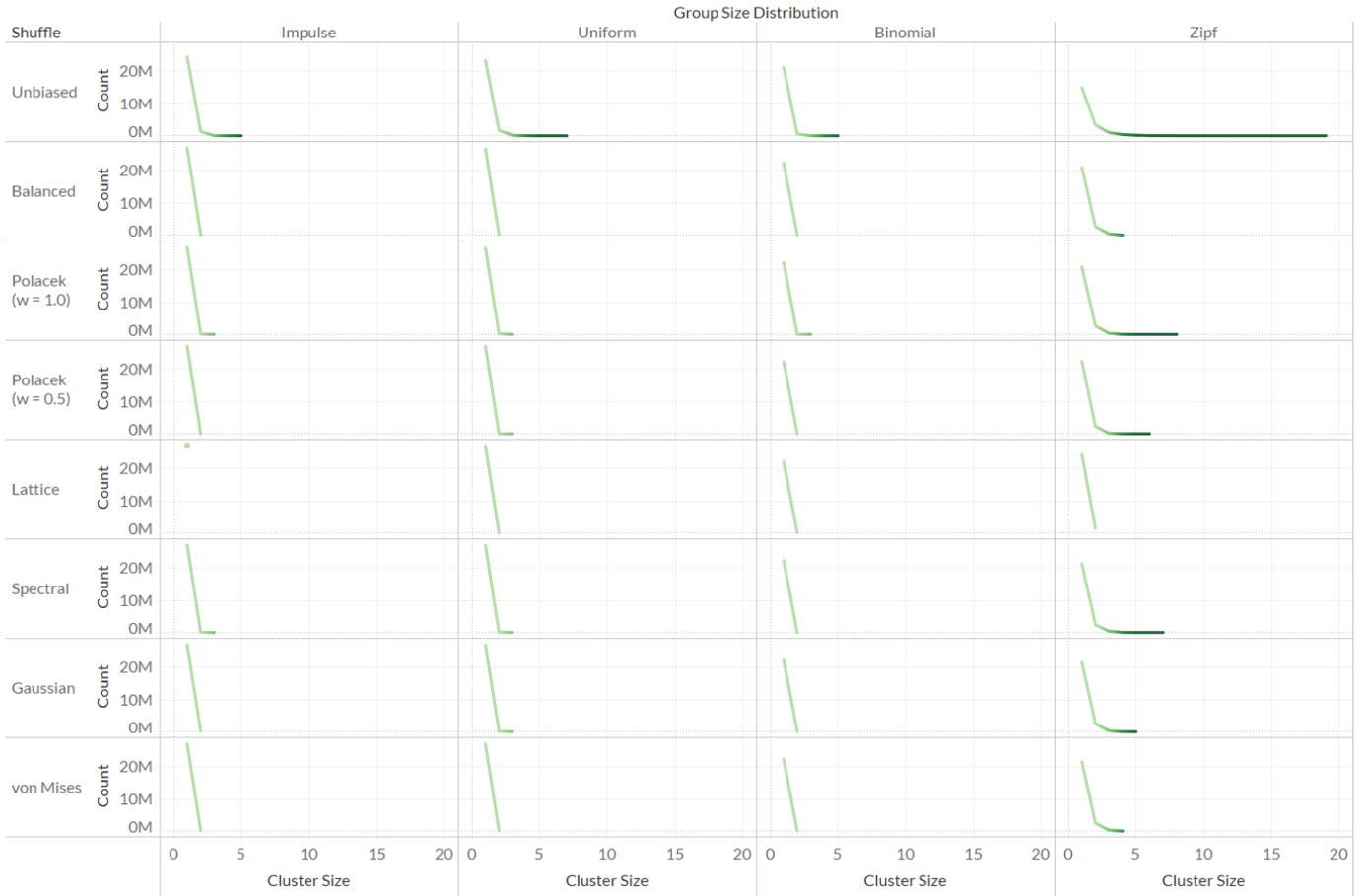

*Figure 37: Cluster size distributions for $10^5$ consecutive shuffle pairs on each of the **large** benchmark playlists using each of the maps.*

## Average Cluster Sizes

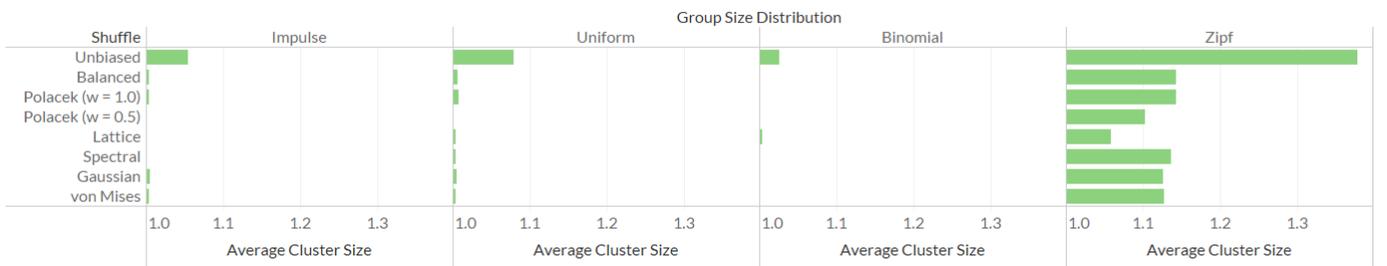

*Figure 38: Average cluster sizes for $10^5$ consecutive shuffle pairs on each of the **large** benchmark playlists using each of the maps.*

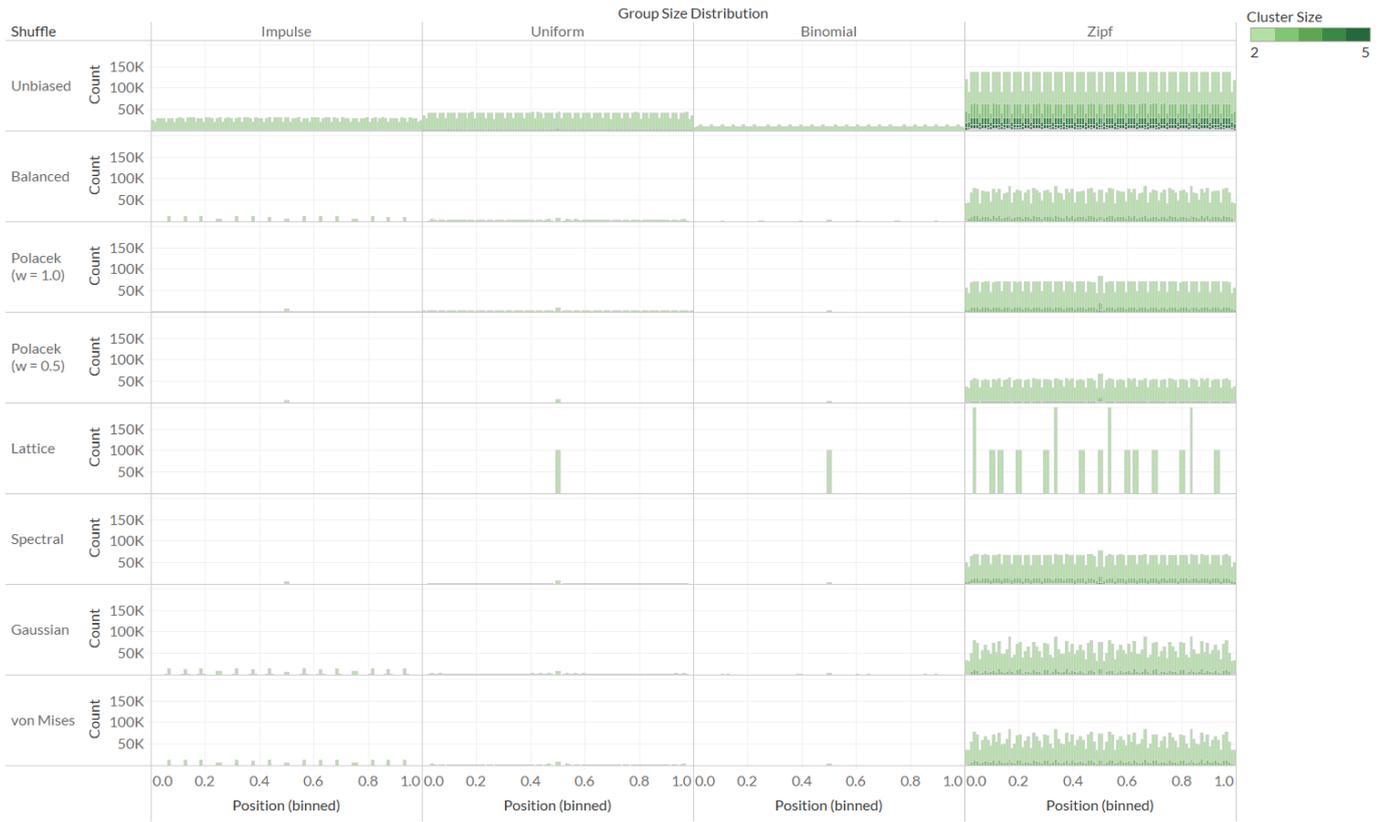

Figure 39: Cluster location distributions for $10^5$ consecutive shuffle pairs on each of the **large** benchmark playlists using each of the maps. Bins are $0.01$ wide.

All of the maps handily outperform unbiased shuffles in cluster size distribution, having smaller maximum cluster sizes and smaller cluster size counts for $2 \leq$ item clusters.

While the lattice map has the best performance in terms of cluster sizes, the location of its clusters are concentrated at specific locations and on the boundary between the two consecutive shuffles, making it clear when we are moving from one shuffle to the next when using shuffle + repeat play.

The addition of a circular shift on top of the Balanced Shuffle, as demonstrated by the Poláček Shuffle with $w = 1.0$, causes an increase in $2 \leq$ item clusters but it also spreads the clusters out more evenly. Decreasing $w$ to $0.5$ is a double-edged sword as well, both simultaneously decreasing average cluster size while also making clusters on the boundary more apparent due to the number of clusters at the boundary staying about the same while decreasing everywhere else.

Between the Spectral, Gaussian, and von Mises maps, the Spectral map has more large clusters compared to the latter two but it also spreads the clusters out more evenly. Between the Gaussian and von Mises maps, the von Mises map has a lower maximum cluster size but shifts more of the clusters towards its maximum cluster size. The von Mises map also tends to spread clusters out more evenly compared to the Gaussian map with smaller differences between the extrema.

In general, we see a trade-off between cluster size and cluster density evenness, with maps producing more large clusters having more consistent cluster density. This comes from large clusters increasing cluster density more uniformly when they are present because they span a wider set of positions than small clusters which only increase cluster density for a small set of positions.

Thus, which map we want to use depends on whether we prioritize cluster sizes or cluster density evenness. In the context of music playback, it may be that having a low maximum cluster size with the larger cluster sizes appearing reasonably often is the preferred map behavior. The second part is key because it shapes the listener's perception of randomness over time. Looking back at the coin flip example, there is strong support for the conjecture stating that people who are asked to emulate a single, short series of fair coin tosses will tend to produce one that is representative of what they think a random process would produce: one with a relatively even number of heads and tails with a high alternation frequency.[4] By keeping maximum cluster size low while still having the larger cluster sizes appear frequently enough, we can make these larger cluster sizes seem less unusual.

Between the different maps we've tested, the Balanced Shuffle's map and the von Mises map behave the closest to this kind of behavior, with the von Mises map producing less large clusters than the Balanced Shuffle's map. Additional testing would be needed to see which one people would prefer or perceive as more random over time.

## 5.3 Alter

Up until now, we've only been concerned with how to interlace item groups together under the assumption that all items in a group are identical. In reality, each item in a group is likely to be different from each other; we wouldn't call Claude Debussy's *Clair de Lune* the same as his *Reflets dans l'eau* just because they're from the same composer. The same goes for any two songs from the same artist, album, genre, or however they're grouped.

Most of the existing shuffling algorithms don't consider shuffle + repeat play scenarios. They use an unbiased shuffle to shuffle individual item groups before mapping and merging them together. This allows a song that was last or near last in the item group to become first or almost first in the next shuffle, letting the song be played twice consecutively or in close succession (Figure 40).

$$1\ 2\ 1\ 4\ 3\ 2\ 1\ 3\ 1\ 2\ |\ 2\ 3\ 2\ 1\ 4\ 1\ 2\ 1\ 3\ 1$$

*Figure 40: An example pair of consecutive shuffles with the same song being played consecutively.*

This is likely to be very distracting to listeners. Like the map step, we'll need to find a way to space songs at relatively even distances from themselves across shuffles without locking them into the same position in their group for every shuffle.

### 5.3.1 Full Alter

First we'll define the full alter, which is simply shuffling each item group with an unbiased shuffle just like the Balanced Shuffle and Poláček Shuffle. This is used for the initial "seed" shuffle in a series of consecutive shuffles (i.e. when shuffle + repeat play is toggled on).

### 5.3.2 Partial Alter

For the partial alter, we'll have each group's items compete to be as close to the middle of the group as possible. The middle items should try to stay near the middle while the outer items should try to jump more positions towards the middle the further out they are. To prevent an item from jumping from one side of the group to the other, we'll impose a maximum jump distance and only allow items to swap once, with self swaps being allowed. Trying to jump to an out-of-bounds position or trying to swap with an item that has already swapped once doesn't count as a swap.

The order items are swapped in is determined by an unbiased shuffle. The later an item gets to swap, the less likely there will be another item for it to swap with because most of the other items will likely have swapped once already. This gives each item a fair chance to swap regardless of their initial position in the group.

To determine which direction and by how many positions an item should jump, we'll use a shifted binomial distribution centered on $0$. The binomial distribution is convenient because it's discrete and its $n$ parameter determines its support, which is on $\mathbb{Z} \in [0, n]$. We'll calculate $n$ using Equation 7, where $n_{group}$ is the number of items in the group and $ceil(x)$ is the ceiling function.

$$n = 2 \cdot ceil(\frac{n_{group} - 1}{4}) \tag{7}$$

This forces $n$ to be even so that a binomial r.v. with this distribution has an odd-length support. Left shifting this binomial r.v. by $ceil(\frac{n_{group}-1}{4})$ moves the support to be on $\mathbb{Z} \in [-ceil(\frac{n_{group}-1}{4}), ceil(\frac{n_{group}-1}{4})]$, centering its middle point on $0$ which allows an item to swap with itself. Defining $n$ this way also makes it so that a song cannot move across more than approximately a quarter of the group in either direction for most group sizes while still allowing for items to swap in $2$ item groups. For larger groups, this prevents a song that was played last or near last in one shuffle to be played first or almost first in the next shuffle.

We can also control the location of the binomial distribution's peak using the $p$ parameter, where decreasing $p$ moves the peak to the left closer to $-ceil(\frac{n_{group}-1}{4})$ post shift, increasing $p$ moves the peak to the right closer to $ceil(\frac{n_{group}-1}{4})$ post shift, and having $p = 0.5$ places the peak in the middle right on $0$ post shift. We'll calculate $p$ using an equation with rotational symmetry of order $2$ around the point $(\frac{n_{group}-1}{2}, 0.5)$ whose range is $[0, 1]$ when the input is on $[0, n_{group} - 1]$. We'll be using a linear equation given by Equation 8, where $i$ is the index of the $i$-th item in a zero-indexed group with $n_{group}$ items.

$$p(i) = \frac{n_{group} - 1 - i}{n_{group} - 1} \tag{8}$$

This makes it so items near the middle of the group have a $p$ close to $0.5$ while items near the beginning or the end of the group have a $p$ close to $1$ or $0$ respectively, with items equidistant from the middle having $p$ values equidistant from $0.5$.

We'll also make it so that the partial alter behaves like an unbiased shuffle when there are only $2$ items in the group. To do that, we'll first need to find the $p$ values for item $0$ and item $1$ such that when $n = 2 \cdot ceil(\frac{2-1}{4}) = 2$, there is a $50\%$ chance that the items swap. Using $p_{a+b}$ to denote the probability of item $a$ trying to swap with the item $b$ positions from it, we can represent this $50\%$ swap chance with Equation 9.

$$\frac{1}{2} \cdot (p_{0+1} + p_{0-1} \cdot p_{1-1}) + \frac{1}{2} \cdot (p_{1-1} + p_{1+1} \cdot p_{0+1}) = \frac{1}{2} \tag{9}$$

The first set of terms is the probability that item $0$ gets to swap first while the second set of terms is the probability that item $1$ gets to swap first. Since the the equation we're using to calculate $p$ will cause items equidistant from the middle of the group to have $p$ values equidistant from $0.5$, item $1$'s $p$ parameter is equal to $1$ minus item $0$'s $p$ parameter.

$$p_1 = 1 - p_0 \tag{10}$$

Using this, we can rewrite Equation 9 in terms of item $0$'s $p$ parameter as Equation 11.

$$p_0^2 + (1 - p_0)^2 \cdot p_0^2 = \frac{1}{2} \tag{11}$$

Solving for $p_0$ and taking the positive real root gives us a rather complicated value.

$$p_0 = \frac{1}{2} - \frac{1}{2\sqrt{-1 - \frac{2}{\sqrt[3]{17+3\sqrt{33}}} + \sqrt[3]{17+3\sqrt{33}}}} +$$
$$\frac{1}{2}\sqrt{-\frac{2}{3} + \frac{2}{3\sqrt[3]{17+3\sqrt{33}}} - \frac{1}{3}\sqrt[3]{17+3\sqrt{33}} + 2\sqrt{\frac{3}{-1 - \frac{2}{\sqrt[3]{17+3\sqrt{33}}} + \sqrt[3]{17+3\sqrt{33}}}}} \tag{12}$$
$$\approx 0.67186$$

Now that we have $p_0$, we want to modify the $p$ calculation in Equation 8 so that it's bounded by $[1 - p_0, p_0]$ while maintaining its rotational symmetry. Renaming $p_1 = 1 - p_0$ to $p_{margin}$, we get our new $p$ calculation given by Equation 13.

$$p(i) = p_{margin} + (1 - 2 \cdot p_{margin}) \cdot \left(\frac{n_{group} - 1 - i}{n_{group} - 1}\right) \tag{13}$$

Overall, the partial alter takes $O(n)$ time and $O(n)$ space.

### 5.3.3 Benchmarks

To check that the partial alter works like a full alter (i.e. an unbiased shuffle) when $n = 2$, we'll look at the distribution of indices each item occupies (the item index distribution) and the distribution of index changes between partial alters for each item (the item index change distribution) (Figures 41, 42). We'll also do the same for $n = 10$ and $n = 100$ to get a more general picture of what the partial alter does (Figures 43, 44, 45, 46).

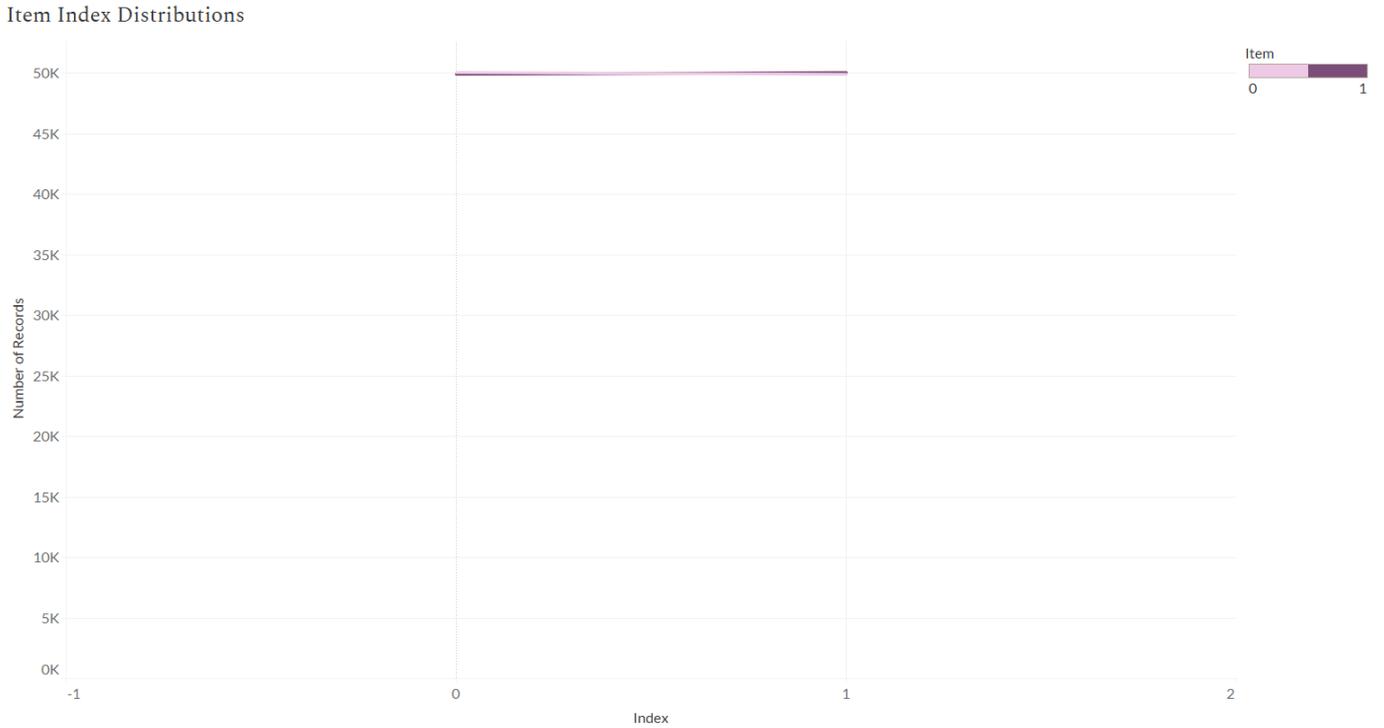

Figure 41: *Item index distributions on a group of size $2$ across $10^5$ partial alters.*

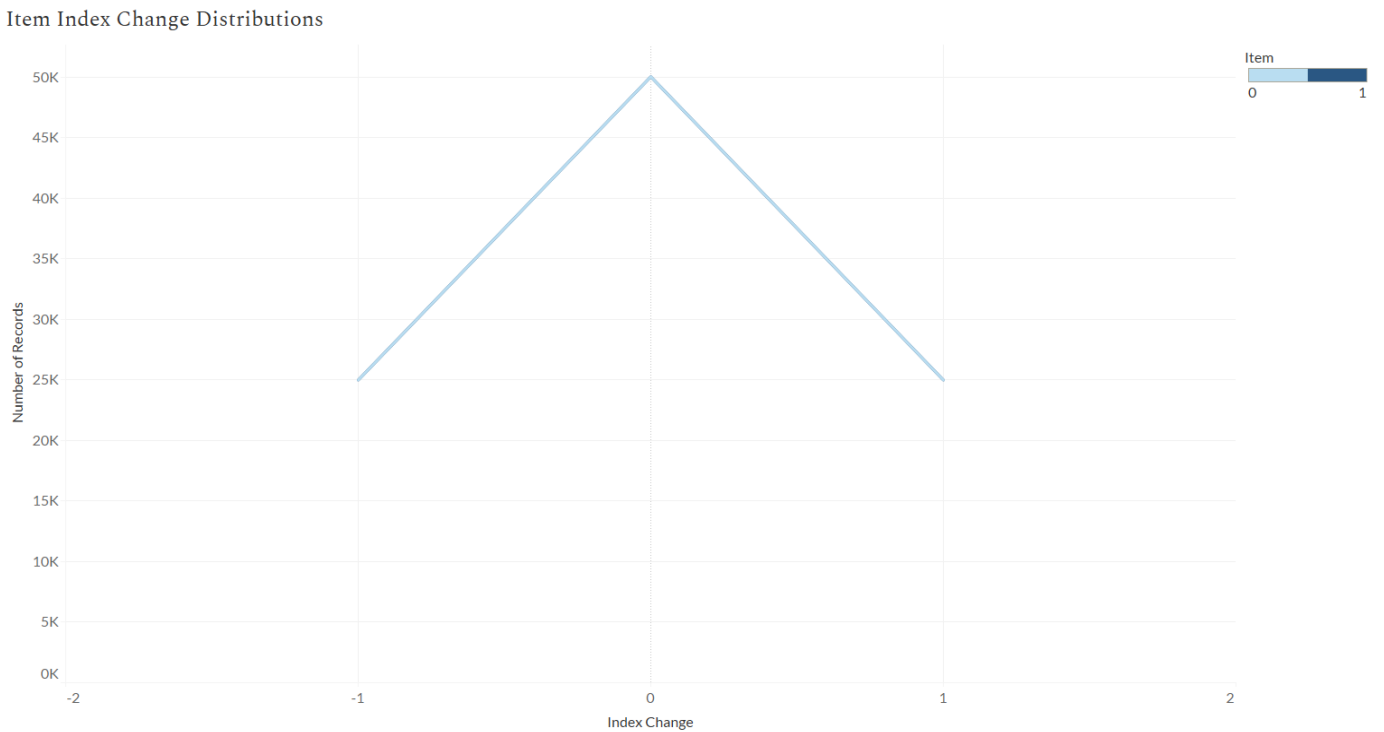

Figure 42: *Item index change distributions on a group of size $2$ across $10^5$ partial alters.*

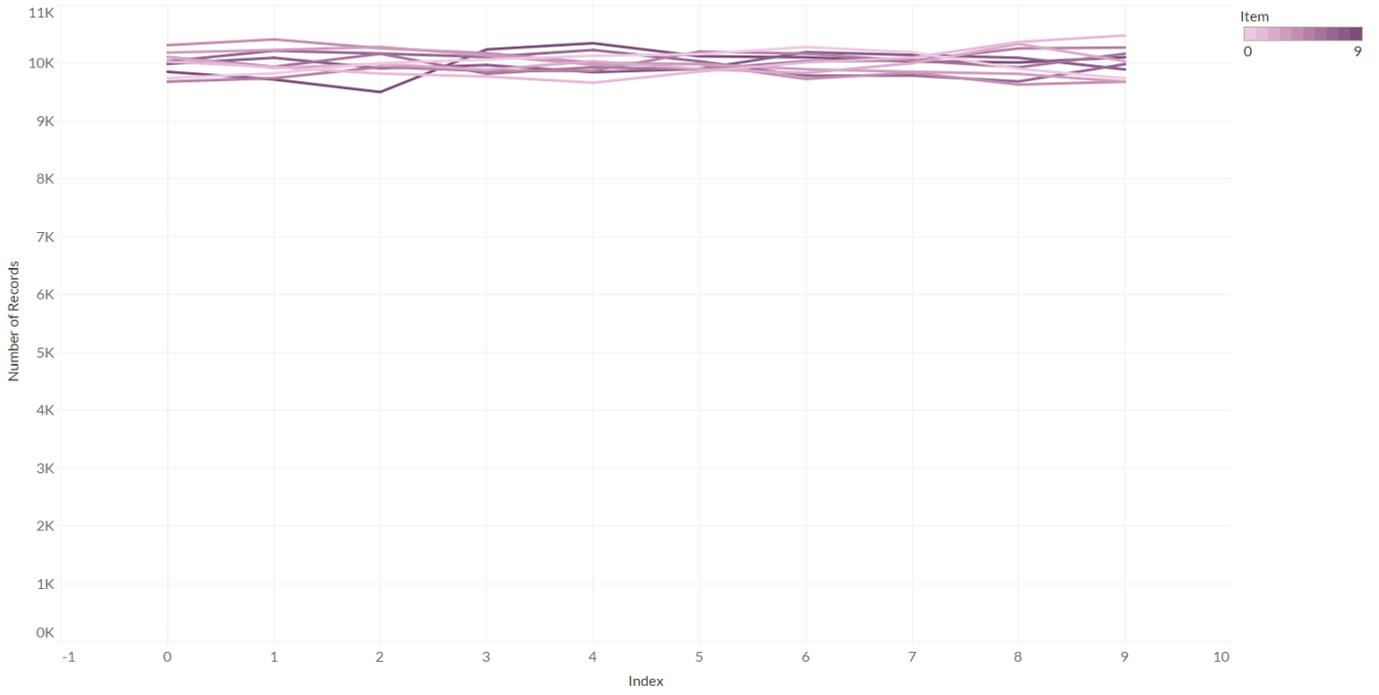

*Figure 43: Item index distributions on a group of size $10$ across $10^5$ partial alters.*

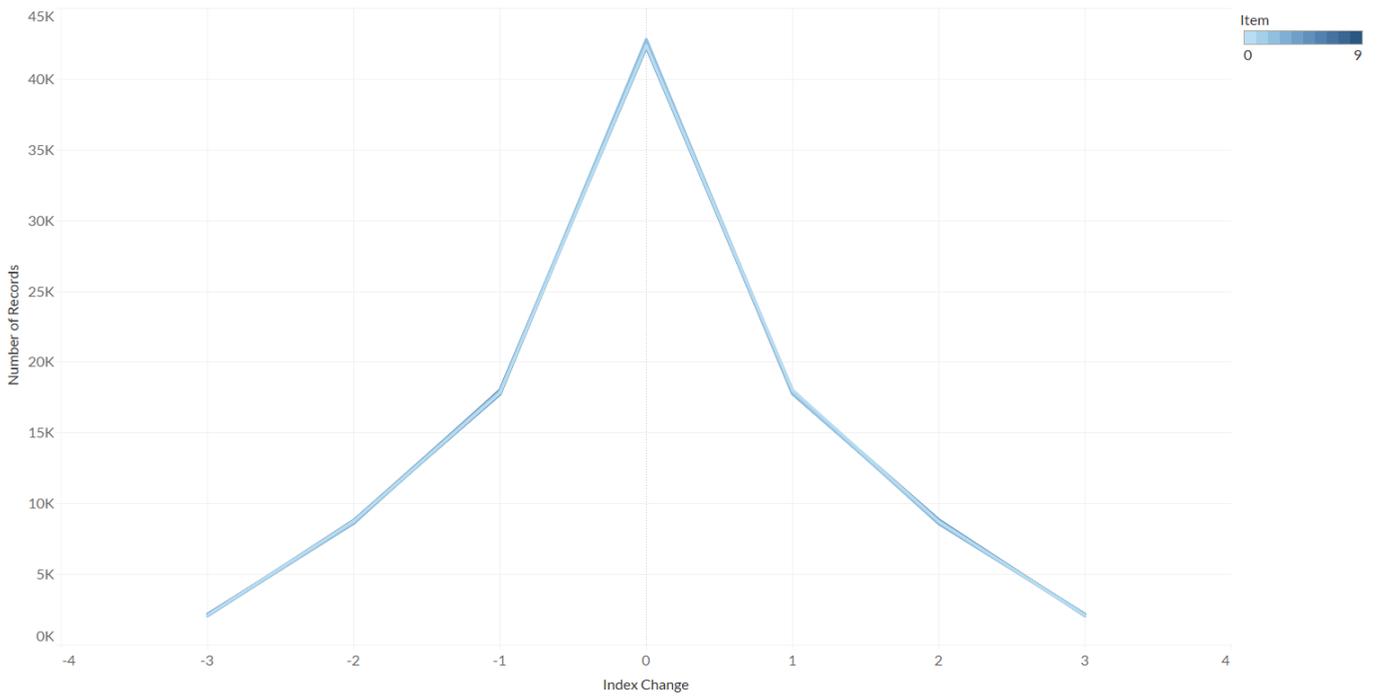

*Figure 44: Item index change distributions on a group of size $10$ across $10^5$ partial alters.*

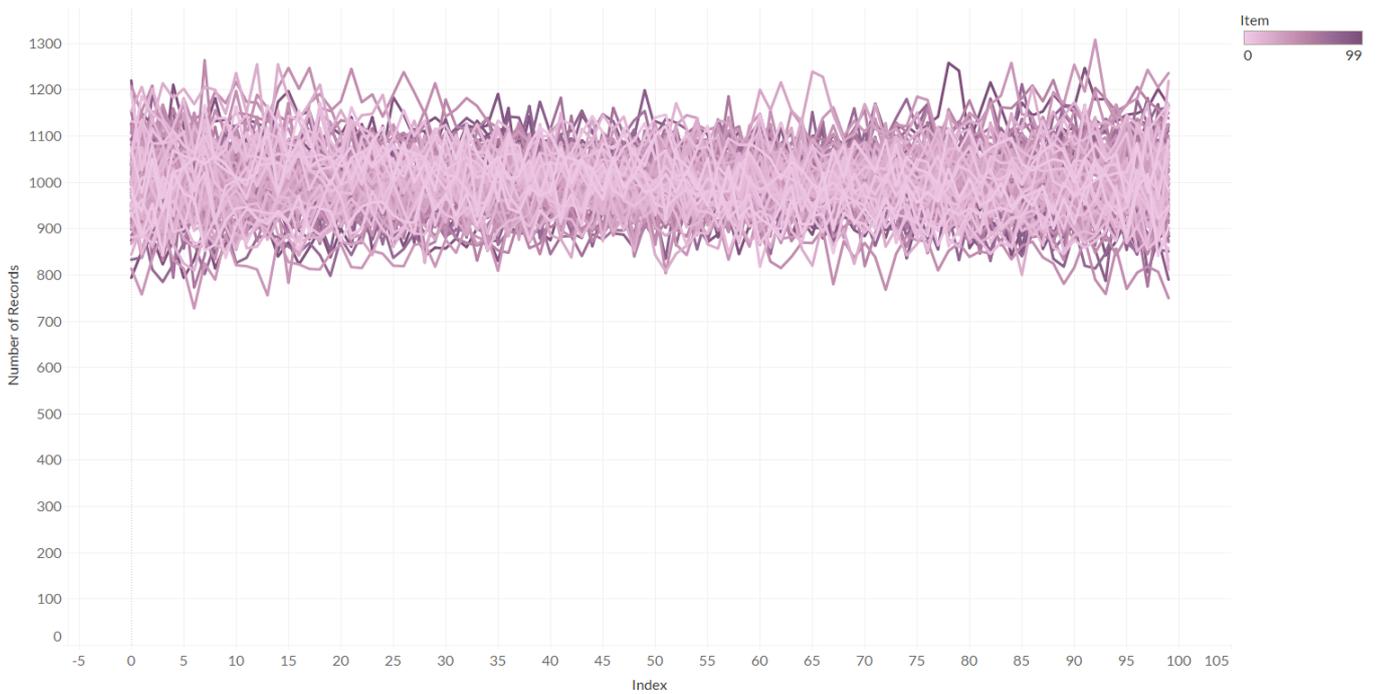

*Figure 45: Item index distributions on a group of size $100$ across $10^5$ partial alters.*

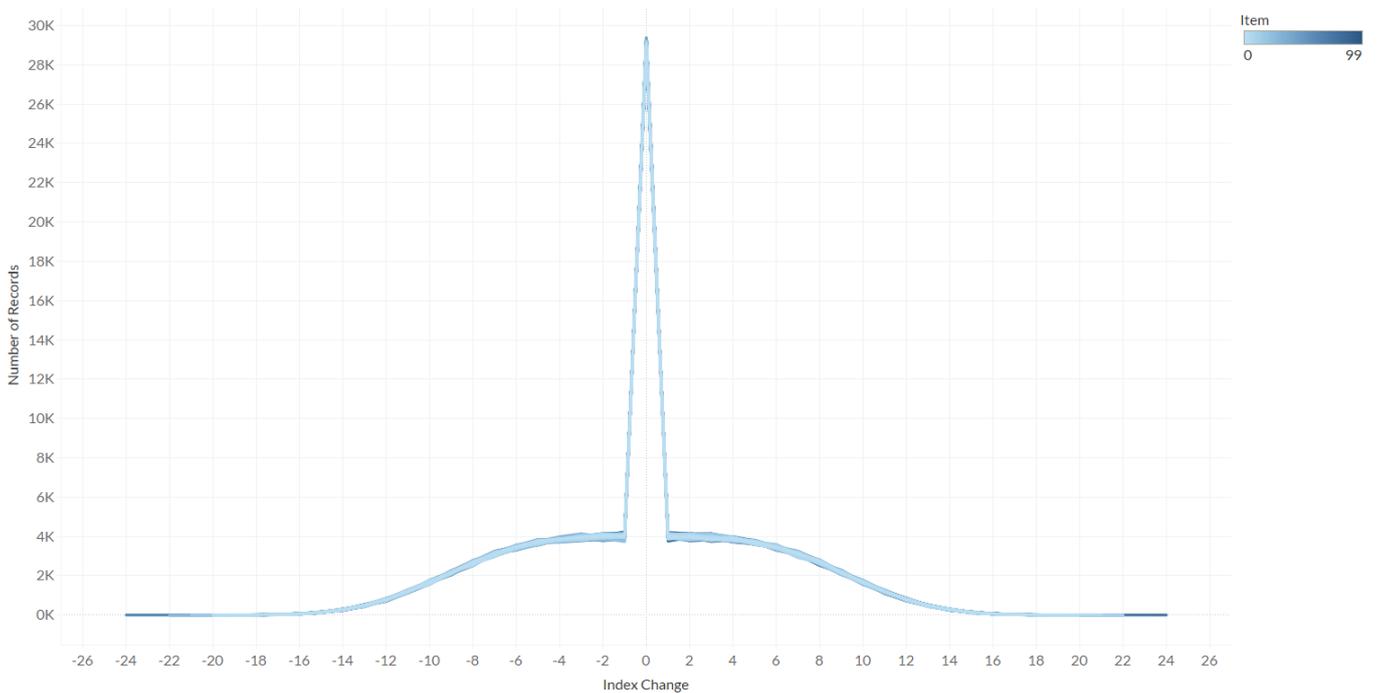

*Figure 46: Item index change distributions on a group of size $100$ across $10^5$ partial alters.*

As desired, the partial alter behaves like a full alter when $n = 2$, with both items being equally likely to be at either index while having a $50\%$ chance to keep the same index and a $25\%$ chance each to move left and right. More interestingly is when $n = 10$ and $100$ where each item is still equally likely to be at any index while the index change distribution looks like a bell-shaped distribution with a spike at $0$, a result of items that try to swap later being less likely to find an item to swap with. Visually, we can see that the partial alter prevents items from jumping too far across a group between consecutive shuffles (Figures 47, 48, 49, 50).

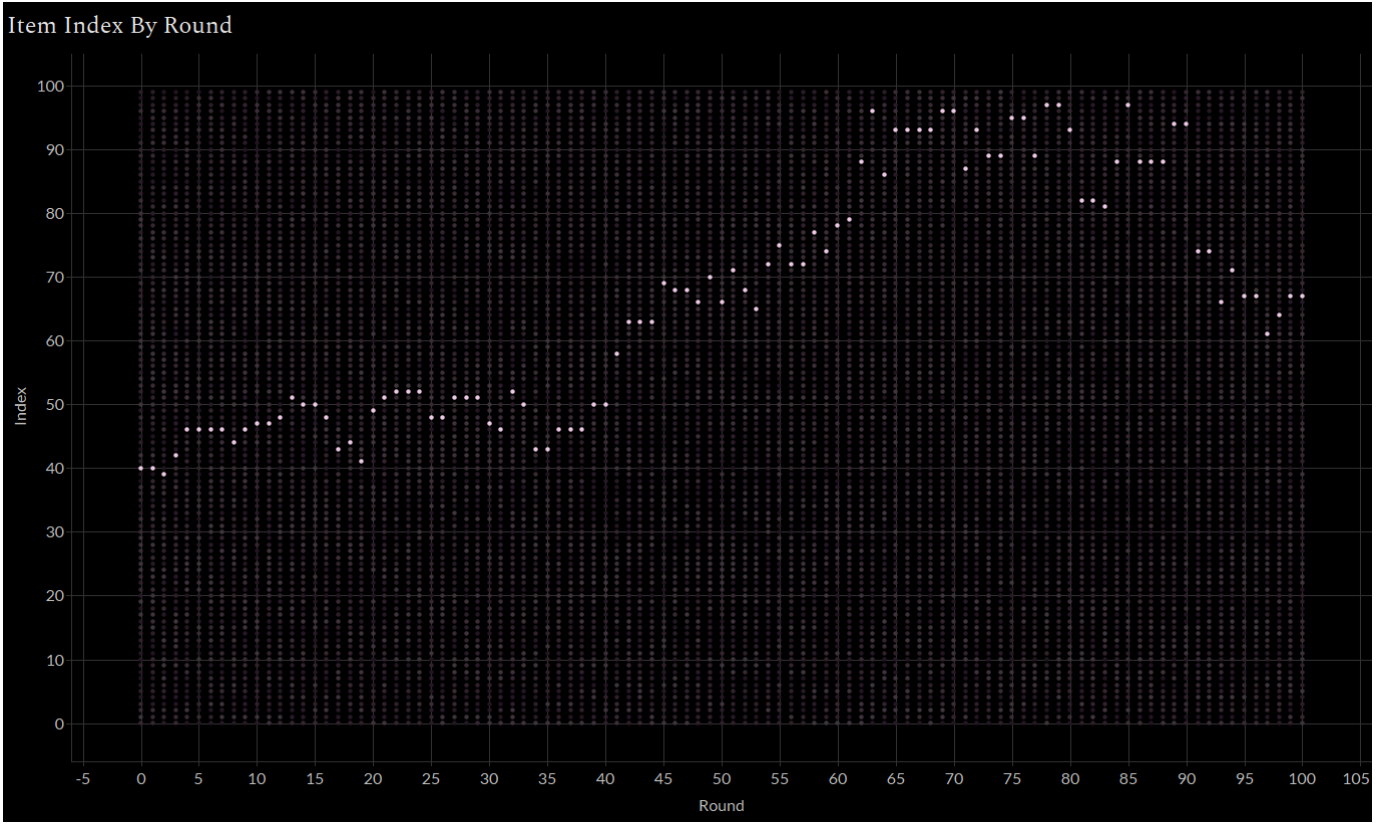

*Figure 47: Item* 0*'s index in a group of size* 100 *across* 100 *partial alters.*

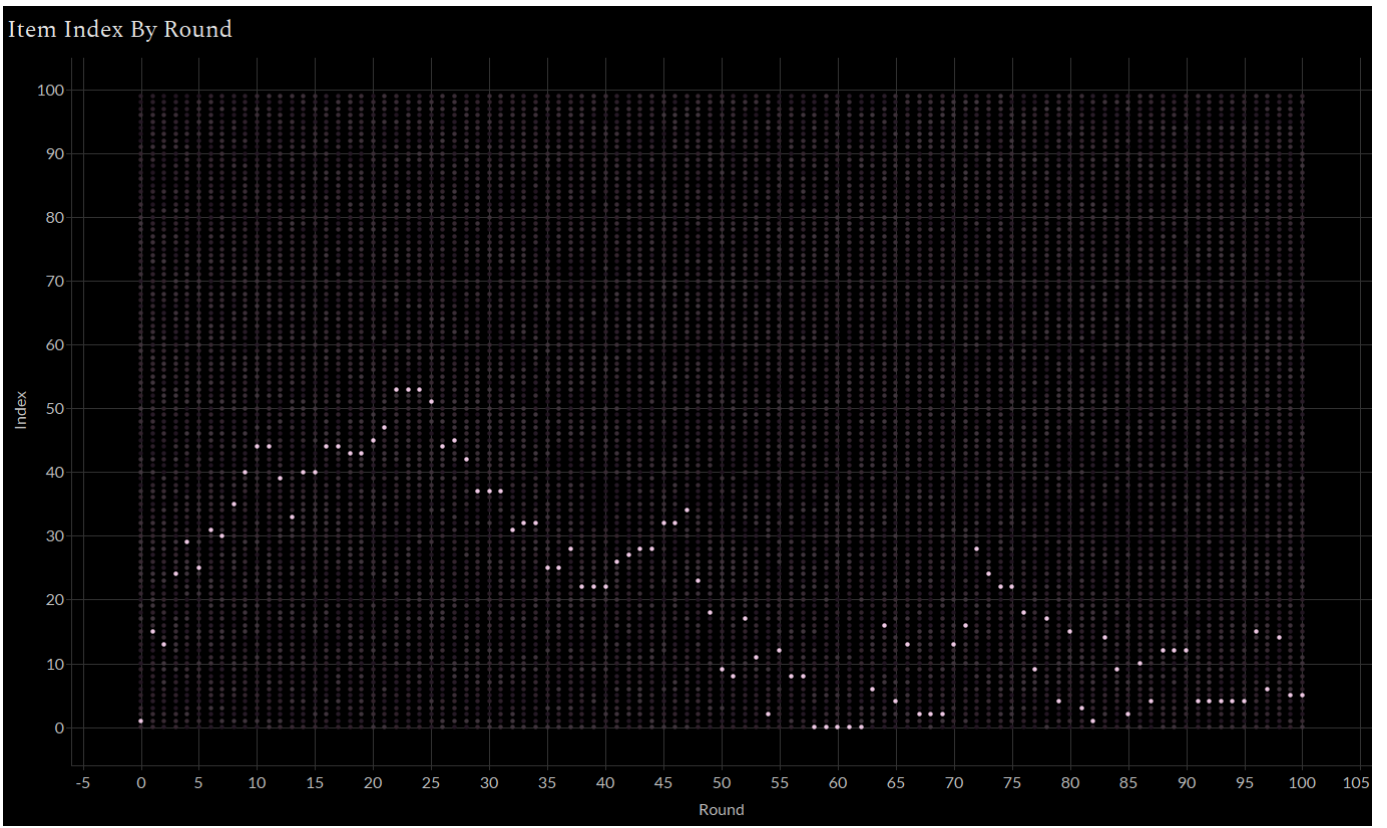

*Figure 48: Item* 1*'s index in a group of size* 100 *across* 100 *partial alters.*

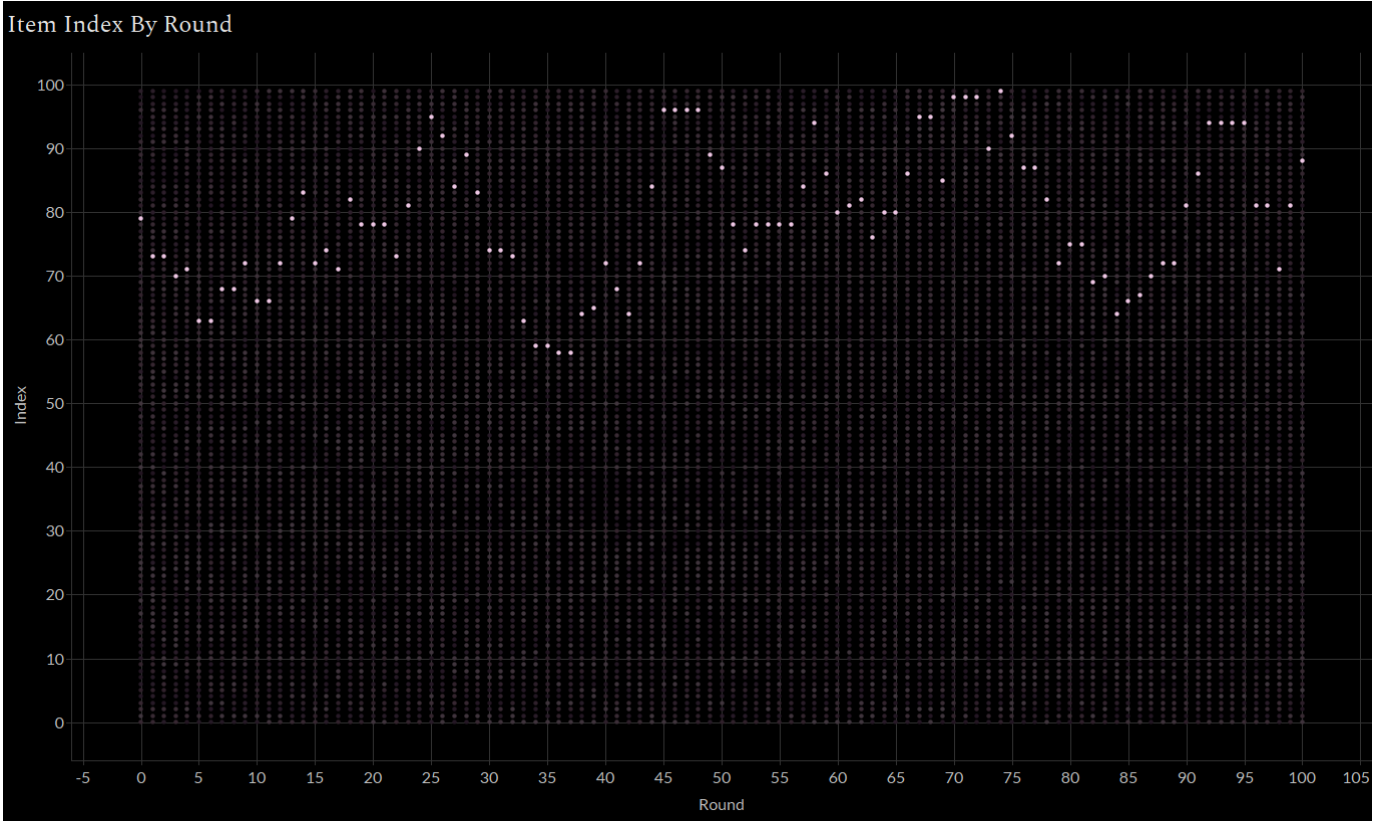

*Figure 49: Item 2's index in a group of size 100 across 100 partial alters.*

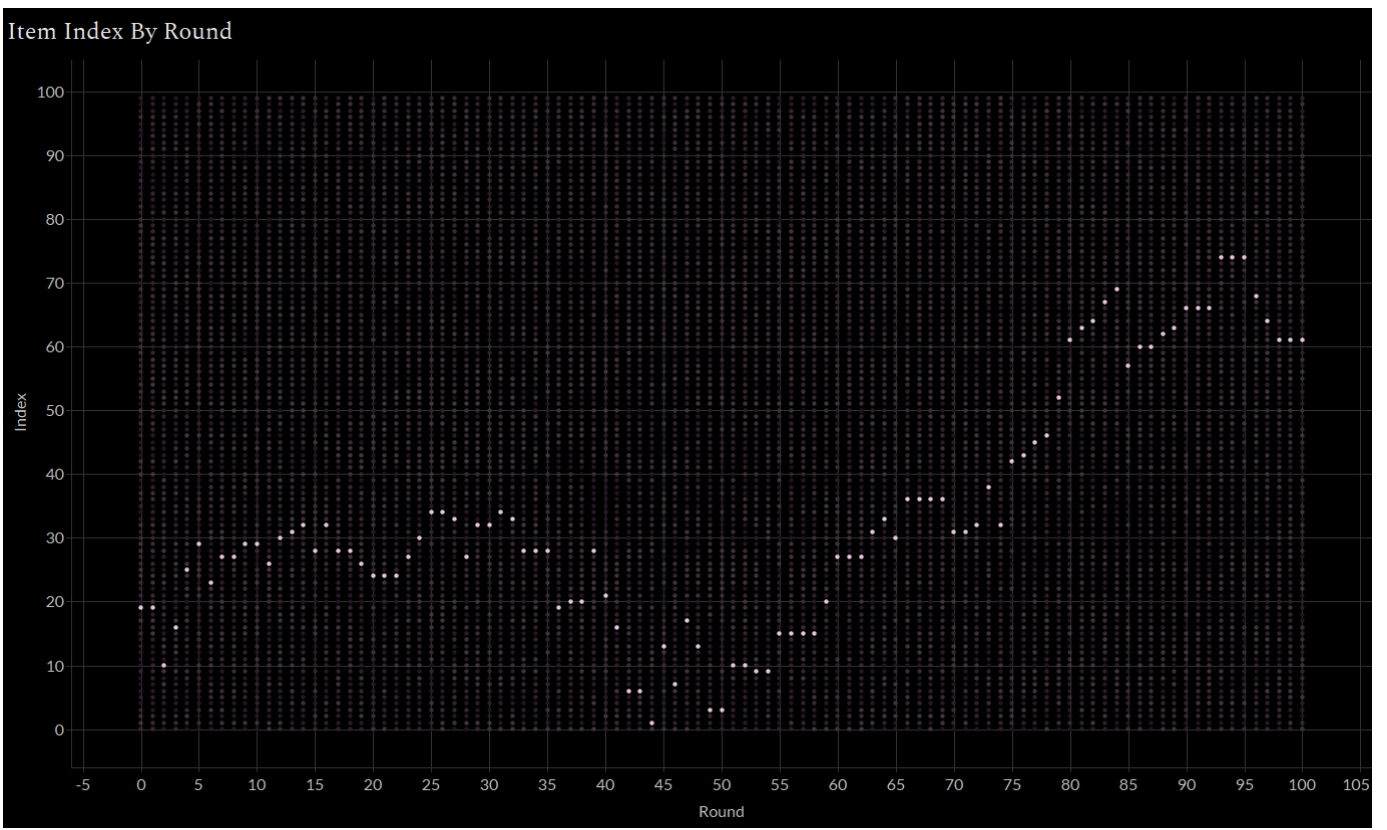

*Figure 50: Item 3's index in a group of size 100 across 100 partial alters.*

# 6 Conclusion

With the cluster diffusing (CD) shuffles, we were able to create a general framework for the different steps of a shuffling algorithm along with various implementations of these alter, map, and merge steps. Whether using the full or partial alter, the combination of the von Mises map or Balanced Shuffle's map with a simple concatenation and radix sort for the merge step allows us to create multi-hyperuniform shuffle play sequences in $O(n)$ time and space, far faster than the Gaussian Ensemble approach which took $O(n^3)$ time and $O(n^2)$ space.

While this paper's contributions end here, further research can be done with regards to the human perception of randomness. The different maps and the partial alter in this paper were all designed with the assumption that disordered hyperuniform and disordered multi-hyperuniform systems appear the most random to humans. Additional research to test this assumption would be interesting and useful for a variety of fields such as psychology, blue noise sampling, and image dithering.

On the topic of blue noise sampling, it also might be interesting to look at using the eigenvalues of matrices from the Complex Ginibre Ensemble (CGE) for two dimensional blue noise sampling and the development of faster approximations of this rather slow method. The CGE is the set of $n \times n$ matrices with i.i.d. complex Gaussian entries. When the entries are normalized to have mean $0$ and variance $\frac{1}{n}$, the eigenvalue distributions of matrices from this ensemble converge to a uniform distribution over the complex unit disk as stated by the circular law (Figure 51).[20]

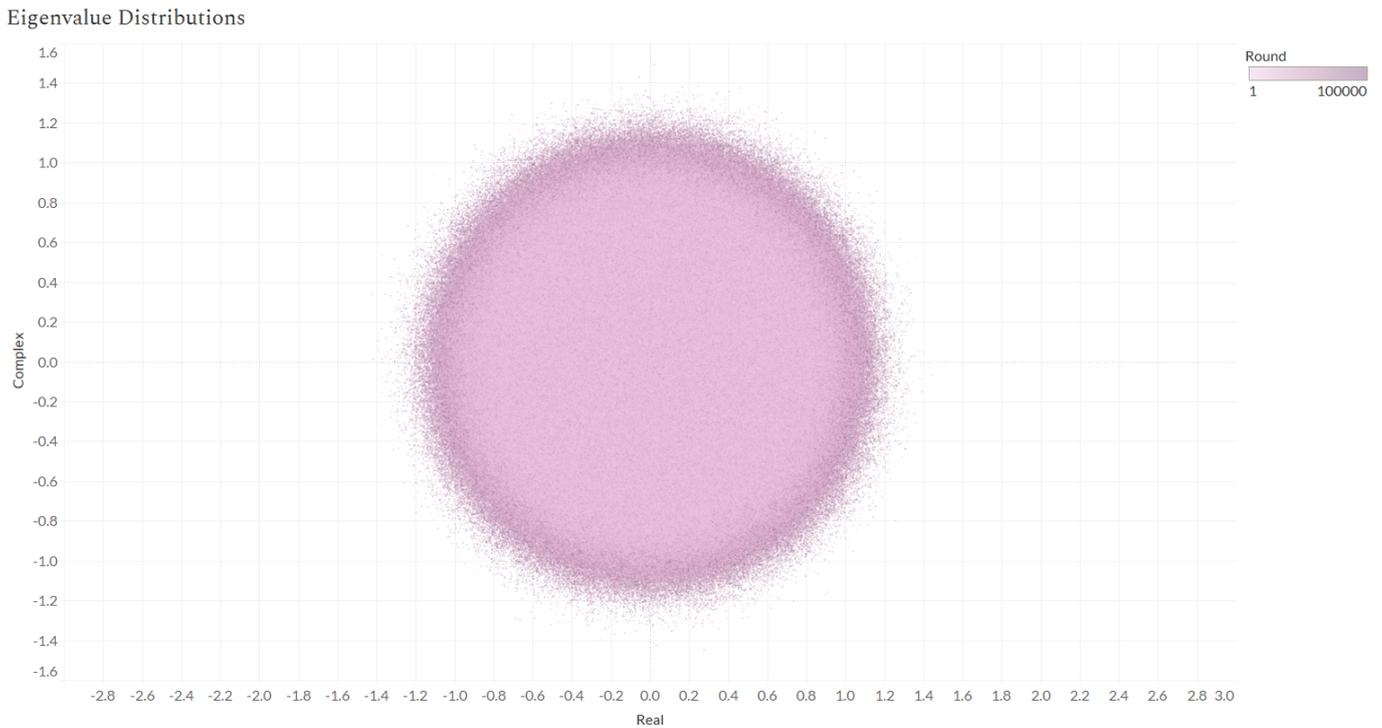

*Figure 51: Eigenvalue distributions for $10^5$ samples of $20 \times 20$ matrices from the CGE.*

The CGE and its sister ensembles have been studied in great detail by many mathematicians and physicists who have shown that their eigenvalues tend to repel each other much like the Gaussian Ensembles and other families of random matrices (Figure 52).[21]

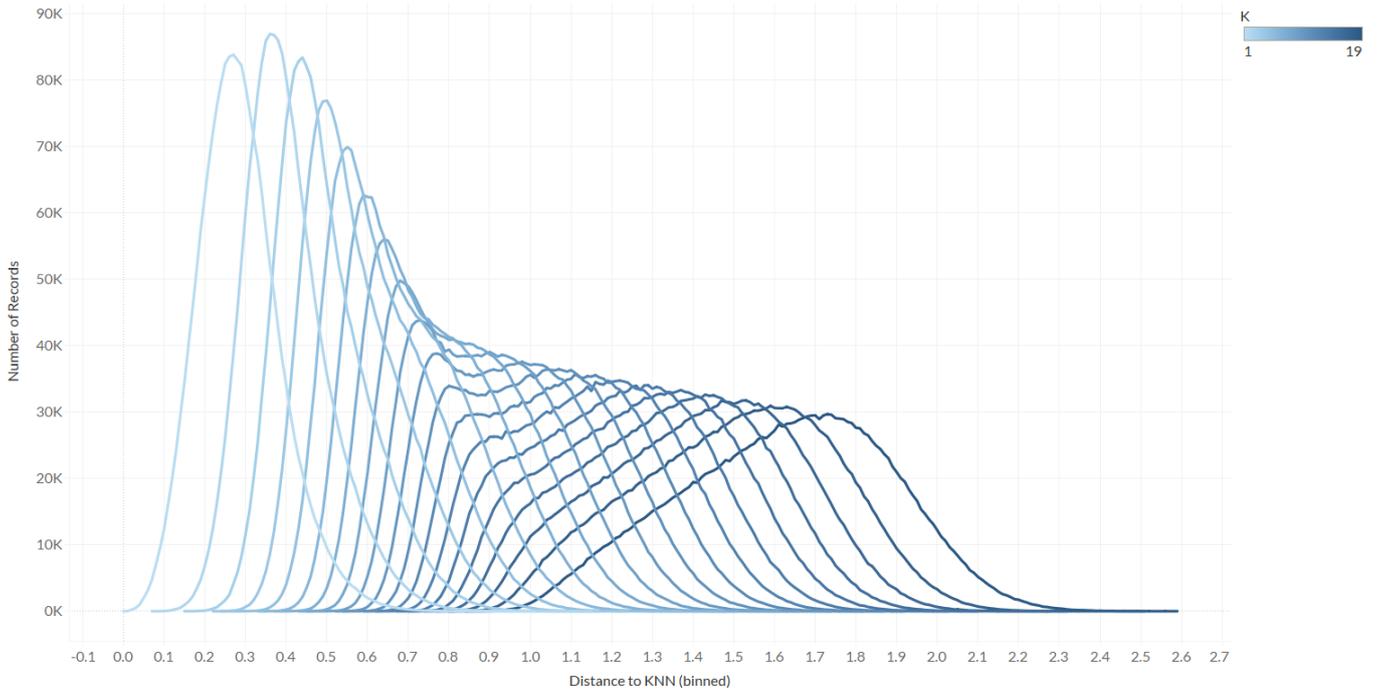

*Figure 52: Eigenvalue $k$-th nearest neighbor spacing distributions for $10^5$ samples of $20 \times 20$ matrices from the CGE.*

If disordered hyperuniform and disordered multi-hyperuniform distributions are perceived to be the most random distributions to humans, it may be worthwhile to look into using the CGE for problems whose aim is to produce two dimensional outputs that appear random to humans.

On a more practical level, research could be done on how to group the input items in different usage scenarios. In the context of music playback, for example, music players may want to group songs on multiple levels and use a CD shuffle recursively on each level. Songs might be grouped by artist at the highest level, much like what Spotify did with the Poláček Shuffle, while also being sub-grouped by album or year. Groups with only one song in the playlist might also be combined into a single larger group to prevent these items from being shuffled in an unbiased manner relative to each other, avoiding the inter-shuffle clusters the partial alter was designed to suppress.